\documentclass[journal,a4paper]{IEEEtran}

\usepackage{cite}
\usepackage{array}
\usepackage{fixltx2e}
\usepackage{color}
\usepackage{psfrag}
\usepackage{epsfig}
\usepackage{tabularx}
\usepackage{amsmath}
\usepackage{amssymb}
\usepackage{amsfonts}
\usepackage{algorithm,algorithmic}
\usepackage{pstricks, pst-node, pst-plot, pst-circ}
\usepackage{moredefs}

\def\PSNR{\mathrm{ PSNR}}

\def\punit{\, \mathrm}

\def\e{\mathrm{ e}}

\def\j{\mathrm{ j}}

\newcommand{\argmin}{\mathop{\mathrm{argmin}}}
\newcommand{\argmax}{\mathop{\mathrm{argmax}}}

\newcolumntype{C}[1]{>{\centering\arraybackslash}p{#1}}

\newcommand{\algorithmicinput}{\textbf{input:}}
\newcommand{\INPUT}{\item[\algorithmicinput]}
\newcommand{\algorithmicoutput}{\textbf{output:}}
\newcommand{\OUTPUT}{\item[\algorithmicoutput]}

\newcolumntype{C}[1]{>{\centering\arraybackslash}p{#1}}

\begin{document}


\title{A Fast Algorithm for Selective Signal Extrapolation with Arbitrary Basis Functions}
\author{J{\"u}rgen~Seiler and~Andr{\'e}~Kaup
\thanks{J. Seiler and A. Kaup are with the Chair of Multimedia Communications and Signal Processing, University of Erlangen-Nuremberg, Cauerstr. 7, 91058 Erlangen, Germany (e-mail: seiler@lnt.de; kaup@lnt.de). } \vspace{-0.5cm}}

\markboth{}%
{Seiler and Kaup: Fast Selective Extrapolation}

\maketitle


\begin{abstract} \label{abstract}
Signal extrapolation is an important task in digital signal processing for extending known signals into unknown areas. The Selective Extrapolation is a very effective algorithm to achieve this. Thereby, the extrapolation is obtained by generating a model of the signal to be extrapolated as weighted superposition of basis functions. Unfortunately, this algorithm is computationally very expensive and, up to now, efficient implementations exist only for basis function sets that emanate from discrete transforms. Within the scope of this contribution, a novel efficient solution for Selective Extrapolation is presented for utilization with arbitrary basis functions. The proposed algorithm mathematically behaves identically to the original Selective Extrapolation, but is several decades faster. Furthermore, it is able to outperform existent fast transform domain algorithms which are limited to basis function sets that belong to the corresponding transform. With that, the novel algorithm allows for an efficient use of arbitrary basis functions, even if they are only numerically defined.

\end{abstract}


\section{Introduction} \label{sec:introduction}

 \IEEEPARstart{T}{he} extrapolation of signals is a very important area in digital signal processing, especially in image and video signal processing. Thereby, unknown or not accessible samples are estimated from known surrounding samples. In image and video processing, signal extrapolation tasks arise e.\ g.\ in the area of concealment of transmission errors as described in \cite{Stockhammer2005} or for prediction in hybrid video coding as shown in \cite{Richardson2003}. 
 
In general, signal extrapolation can be regarded as underdetermined problem as there are infinitely many different solutions for the signal to be estimated, based on the known samples. According to \cite{Olshausen1997}, sparsity-based algorithms are well suited for solving underdetermined problems as these algorithms are able to cover important signal characteristics, even if the underlying problem is underdetermined. These algorithms can be applied well to image and video signals, as in general natural signals are sparse \cite{Candes2007} in certain domains, meaning that they can be described by only few coefficients.
 
As has been shown in \cite{Tropp2004,Temlyakov2000}, out of the group of sparse algorithms the greedy sparse algorithms are of interest, as these algorithms are able to robustly solve the problem. One algorithm out of this group is e.\ g.\ Matching Pursuits from \cite{Mallat1993}. Another powerful greedy sparse algorithm is the Selective Extrapolation (SE) from \cite{Kaup2005}. SE iteratively generates a model of the signal to be extrapolated as weighted superposition of basis functions. In the past years, this extrapolation algorithm also has been adopted by several others like \cite{Herraiz2008, Friebe2007} to solve extrapolation problems in their contexts. 

Unfortunately, SE as it exists up to now is computationally very expensive. This holds except for the case that basis function sets are regarded that emanate from discrete transforms. In such a case, the algorithm can be efficiently carried out in the transform domain. The functions of the Discrete Fourier Transform (DFT) \cite{Cooley1969} are one example for such a basis function set. Using this set, an efficient implementation in the Fourier domain exists by Frequency Selective Extrapolation (FSE) \cite{Kaup2005}. If basis function sets are regarded that do not emanate from discrete transforms or overcomplete basis function sets or even only numerically defined basis functions, such transform domain algorithms cannot exist. 

\begin{figure}
	\begin{center}
	\psfrag{Original}[c][c][0.85]{Original}
	\psfrag{Distorted}[c][c][0.85]{Distorted}
	\psfrag{Selective Extrapolation}[c][c][0.85]{SE with}
	\psfrag{with Fourier basis}[c][c][0.85]{Fourier basis}
	\psfrag{functions}[c][c][0.85]{functions}
	\psfrag{with alternative}[c][c][0.85]{alternative basis}
	\psfrag{basis functions}[c][c][0.85]{functions}
	\psfrag{PSNR=21.58dB}[c][c][0.75]{$\PSNR=\!21.58\punit{dB}$}
	\psfrag{PSNR=24.43dB}[c][c][0.75]{$\PSNR=\!24.43\punit{dB}$}
	\psfrag{PSNR=22.30dB}[c][c][0.75]{$\PSNR=\!22.30\punit{dB}$}
	\psfrag{PSNR=23.27dB}[c][c][0.75]{$\PSNR=\!23.27\punit{dB}$}
	\psfrag{PSNR=21.82dB}[c][c][0.75]{$\PSNR=\!21.82\punit{dB}$}
	\psfrag{PSNR=32.46dB}[c][c][0.75]{$\PSNR=\!32.46\punit{dB}$}
	\includegraphics[width=0.485\textwidth]{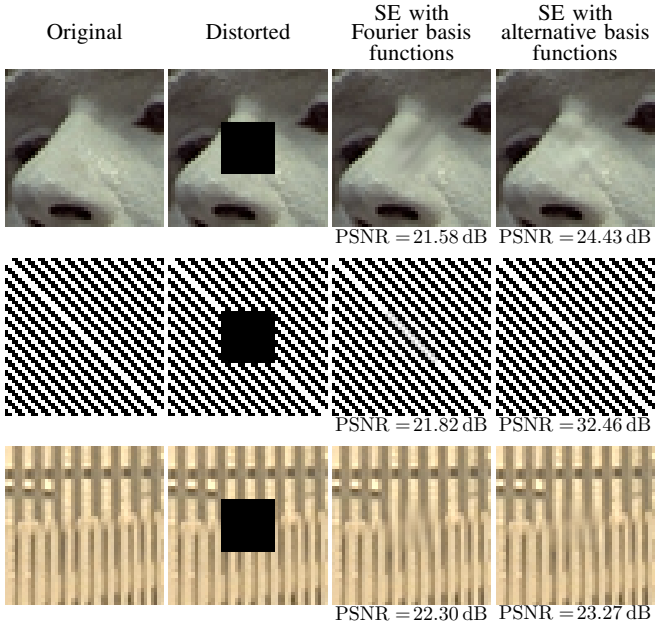}
	\caption{Examples for image signals where Fourier basis function provide insufficient extrapolation quality. In every row, original signal, distorted signal, and extrapolated signals are shown. Extrapolation is carried out either with DFT basis functions or alternative ones. Top row: union of DCT and WHT basis functions. Mid row: binarized DFT basis functions. Bottom row: union of DFT and binarized DFT basis functions.\vspace{-4mm}}
	\label{fig:bf_examples}
	\end{center}
	
\end{figure}

Although Fourier basis functions have proven to form a good set for a wide range of signals, there also exist signals where other basis function sets lead to better extrapolation results. This holds for example for the case that the support area on which the extrapolation is based is very unequal or in the case that very steep signal changes occur as e.\ g.\ in artificial signals. Fig.\ \ref{fig:bf_examples} shows three examples for such signals. The left column shows the original signal, the second column shows a distorted signal with the area to be extrapolated marked in black. The signals in the third column result from applying FSE which utilizes Fourier basis functions. In the last column, Selective Extrapolation is carried out with different basis function sets. In the first row, the basis function set results from the union of the functions from the Discrete Cosine Transform (DCT) \cite{Ahmed1974} and the Walsh-Hadamard Transform (WHT) \cite{Walsh1923}. In the second row, a binarized version of DFT functions is used in order to reconstruct the steep changes in this artificial signal. In the third row, the basis function set emanates from the union of DFT functions and binarized DFT functions. The three examples have in common that the used basis function sets produce significantly better subjective as well as objective results than the Fourier-based extrapolation does. But they have also in common that for such sets no efficient transform domain implementation can exist which would be necessary for a fast implementation.

Within the scope of this contribution we want to introduce a novel spatial domain solution for SE which is called Fast Selective Extrapolation (FaSE). This algorithm is able to generate a model of the signal for arbitrary basis functions in the same way as the original SE, even in the case that the basis function set does not possess any structure and the basis functions are only numerically defined or in the case that an overcomplete basis function set is regarded. But at the same time, the algorithm is very fast as it can efficiently trade computational complexity versus memory consumption. The paper is organized as follows: first, SE will be reviewed for the general case of complex-valued basis functions. With that, an overview of the algorithm is given and the computationally most expensive steps are pointed out. After that, the novel Fast Selective Extrapolation is presented in detail and its complexity is compared to SE. Finally, simulation results are given for proving the abilities of the novel algorithm. 


\section{Review of Selective Extrapolation} \label{sec:se}

\begin{figure}
	\centering
	\psfrag{m}[c][c]{$m$}
	\psfrag{n}[c][c]{$n$}
	\psfrag{Loss area}[l][l]{Loss area $\mathcal{B}$}
	\psfrag{Support area}[l][l]{Support area $\mathcal{A}$}
	\includegraphics[width=0.225\textwidth]{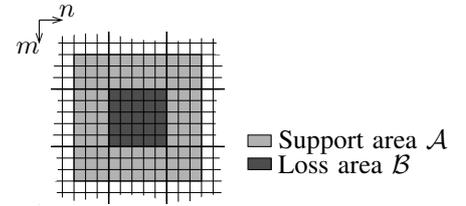}\vspace{-1mm}
	\caption{Extrapolation area $\mathcal{L}$ consisting of loss area $\mathcal{B}$ and support area $\mathcal{A}$.\vspace{-4mm}}
	\label{fig:extrapolation_area}
\end{figure}

For the presentation of Selective Extrapolation (SE) a scenario as shown in Fig.\ \ref{fig:extrapolation_area} is regarded. There, signal parts which have to be extrapolated are subsumed in loss area $\mathcal{B}$. For extrapolating the signal, surrounding correctly received signal parts are used. These signal parts form the support area $\mathcal{A}$. The two areas together form the so called extrapolation area $\mathcal{L}$ which is of size $M\times N$ samples and is depicted by the spatial coordinates $m$ and $n$. The signal in $\mathcal{L}$ is denoted by $s\left[m,n\right]$, but is only available in the support area $\mathcal{A}$. The extrapolation of square blocks is used for presentational reasons at this point only. In general, arbitrarily shaped regions can be extrapolated. In addition to that, in general, the used basis functions can as well be larger than the regarded extrapolation area. In such a case, the extrapolation area has to be padded with zeros to be of the same size as the basis functions. But, for presentational reasons we also assume that the extrapolation area and the basis functions have the same size subsequently. 

As described in \cite{Kaup2005}, SE aims at generating a parametric model $g\left[m,n\right]$ for signal $s\left[m,n\right]$ in whole area $\mathcal{L}$. The model 
 \begin{equation}
 \label{eq:model} g\left[m,n\right] = \sum_{k\in\mathfrak{K}} \hat{c}_k\varphi_k\left[m,n\right]
\end{equation}
emanates from a weighted superposition of the basis functions $\varphi_k\left[m,n\right]$ which are defined over complete $\mathcal{L}$ and are indexed by $k$. Set $\mathfrak{K}$ contains the indices of all basis functions used for model generation. As not all possible basis functions are used for the model, set $\mathfrak{K}$ is a subset of dictionary $\mathfrak{D}$ which holds all basis functions. In order to control the weights of the individual basis functions, one expansion coefficient $\hat{c}_k$ is assigned to each basis function $\varphi_k\left[m,n\right]$. The challenge is to determine which basis functions to use for the model and to calculate the corresponding weights. SE solves this problem iteratively, at which in every iteration one basis function is selected and the corresponding weight is estimated. This is achieved by successively approximating signal $s\left[m,n\right]$ in support area $\mathcal{A}$ and identifying the dominant basis functions of the signal. In doing so, the signal can be continued well into area $\mathcal{B}$, if an appropriate set of basis functions is used.

Initially, model $g^{\left(0\right)}\left[m,n\right]$ is set to zero and with that the initial approximation residual 
\begin{equation}
r^{\left(0\right)}\left[m,n\right] = s\left[m,n\right]
\end{equation}
is equal to the original signal. At the beginning of each iteration, in general the $\nu$-th iteration, a weighted projection of the residual onto each basis function is conducted. For every basis function, this leads to the projection coefficient
\begin{equation}
\label{eq:projection_complex}
p_k^{\left(\nu\right)} = \frac{\displaystyle \sum_{\left(m,n\right)\in\mathcal{L}} r^{\left(\nu-1\right)} \left[m,n\right] \varphi^\ast_k\left[m,n\right] w\left[m,n\right]}{\displaystyle \sum_{\left(m,n\right)\in\mathcal{L}} \varphi_k^\ast\left[m,n\right]w\left[m,n\right]\varphi_k\left[m,n\right]},\ \forall k.
\end{equation}
which results from the quotient of the weighted scalar product between the residual and the basis function and the weighted scalar product between the basis function and itself. In this context, the weighting function 
\begin{equation}
 w\left[m,n\right] = \left\{ \begin{array}{ll} \rho\left[m,n\right] & \mbox{for } \left(m,n\right)\in \mathcal{A} \\ 0 & \mbox{for } \left(m,n\right)\in \mathcal{B}\end{array}\right.
\end{equation}
has two tasks. Firstly, it is used to mask area $\mathcal{B}$ from the calculation of the scalar product as there is no information available about the signal. Secondly, using the function $\rho\left[m,n\right]$ it can control the influence different samples have on the model generation depending on their position. For instance, samples far away from loss area $\mathcal{B}$ can get a smaller weight and due to this weaker influence on the model generation compared to the samples close to area $\mathcal{B}$. In \cite{Meisinger2004b}, an exponentially decreasing weight 
\begin{equation}
 \rho\left[m,n\right] = \hat{\rho}^{\sqrt{\left(m-\frac{M-1}{2}\right)^2+\left(n-\frac{N-1}{2}\right)^2}}
\end{equation}
is proposed with $\hat{\rho}$ controlling the decay.

After the projection coefficients have been calculated for all basis functions, one basis function has to be selected to be added to the model in the actual iteration. The choice falls on the basis function that minimizes the weighted distance
\begin{equation}
e_k^{\left(\nu\right)} \hspace{-1mm}=\hspace{-3mm} \sum_{\left(m,n\right)\in\mathcal{L}}\hspace{-1mm} \left(\left|r^{\left(\nu-1\right)}\left[m,n\right] -p_k^{\left(\nu\right)}\varphi_k\left[m,n\right]\right|^2 w\left[m,n\right]\right)
\end{equation}
between the approximation residual $r^{\left(\nu-1\right)}\left[m,n\right]$ and the projection $p_k^{\left(\nu\right)}\varphi_k\left[m,n\right]$ onto the according basis function.  In this process, again weighting function $w\left[m,n\right]$ from above is used. Hence, the index $u^{\left(\nu\right)}$ of the basis function to be added in the $\nu$-th iteration is:
\[
 u^{\left(\nu\right)}  = \argmin_k \left(e_k^{\left(\nu\right)}\right) \hspace{5.5cm}\vspace{-0.3cm}
\]
\begin{equation}
\label{eq:bf_selection_complex}
  = \argmax_k \hspace{-1mm}\left( \left|p_k^{\left(\nu\right) }\right|^2 \hspace{-0.25cm}\sum_{\left(m,n\right)\in\mathcal{L}} \varphi_k^\ast\left[m,n\right]w\left[m,n\right]\varphi_k\left[m,n\right]\right)\hspace{-1mm}.
 \end{equation}

Subsequent to the basis function selection, the corresponding weight has to be determined. In this process it has to be noted, that although the basis functions may have been orthogonal with respect to the complete extrapolation area $\mathcal{L}$ they cannot be anymore if the scalar products are evaluated in combination with the required weighting function. This effect is not considered in the original paper from \cite{Kaup2005} and is called orthogonality deficiency and is described in detail in \cite{Seiler2007}. In \cite{Seiler2008} fast orthogonality deficiency compensation is proposed to efficiently estimate the expansion coefficient by taking only the fraction $\gamma$ of the projection coefficient:
\begin{equation}
\label{eq:fast_od}
 \hat{c}_{u^{\left(\nu\right)}} = \gamma \cdot p_{u^{\left(\nu\right)}}^{\left(\nu\right)}.
\end{equation}
The factor $\gamma$ is between zero and one and depends on the extrapolation scenario, as described in detail in \cite{Seiler2008}.

After one basis function has been selected and the corresponding weight has been determined, the model and the residual have to be updated by adding the selected basis function to the model generated so far:
\begin{equation}
\label{eq:model_update}
  g^{\left(\nu\right)}\left[m,n\right] = g^{\left(\nu-1\right)}\left[m,n\right] + \hat{c}_{u^{\left(\nu\right)}}\varphi_{u^{\left(\nu\right)}}\left[m,n\right]
\end{equation}
The approximation residual can be updated in the same way and results in
\begin{equation}
\label{eq:residual_update}
r^{\left(\nu\right)}\left[m,n\right] = r^{\left(\nu-1\right)}\left[m,n\right] - \hat{c}_{u^{\left(\nu\right)}} \varphi_{u^{\left(\nu\right)}} \left[m,n\right].
\end{equation}

The above described iterations are repeated until the predefined number of $I$ iterations is reached. Finally, area $\mathcal{B}$ is cut out of the model and is used for replacing the lost signal.

Alg.\ \ref{algo:se} shows the pseudo code of SE for giving a compact overview of this algorithm. Regarding this code and taking into account the equations above, the weighted projection onto all the basis functions in every iteration can be identified as computationally most expensive step. To obtain the projection, a weighted scalar product between the residual and every basis function has to be carried out, leading to a large number of multiplications and additions. Compared to this, the actual basis function selection, the expansion coefficient estimation and the model and residual update have a very small complexity.

\begin{algorithm}[t]
\caption{Selective Extrapolation for arbitrary basis functions}
\fontsize{9}{8}\selectfont
\label{algo:se}
\begin{algorithmic}
\INPUT distorted signal $s\left[m,n\right]$, weighting function $w\left[m,n\right]$, basis functions $\varphi_k\left[m,n\right]$
	\STATE /* Initial residual is equal to original signal */
	\STATE $r\left[m,n\right] = s\left[m,n\right], \forall \left(m,n\right)$
	\FORALL {$\nu=1, \ldots, I$}
	  \STATE /* Projection onto basis functions */
	  \FORALL {$k=0, \ldots, \left|\mathfrak{D}\right|-1$}
		  \STATE $p_k=\frac{\sum_{\left(m,n\right)\in\mathcal{L}} r \left[m,n\right] \varphi^\ast_k\left[m,n\right] w\left[m,n\right]}{\sum_{\left(m,n\right)\in\mathcal{L}} \varphi_k^\ast\left[m,n\right]w\left[m,n\right]\varphi_k\left[m,n\right]}$
	  \ENDFOR	
		\STATE /* Basis function selection */
		\STATE $ u \hspace{-1mm}=\hspace{-1mm} \argmax_k \hspace{-1mm}\left( \left|p_k\hspace{-0.5mm}\right|^2 \hspace{-0.5mm}\sum_{\left(m,n\right)\in\mathcal{L}} \varphi_k^\ast\hspace{-0.5mm}\left[m,n\right]w\hspace{-0.5mm}\left[m,n\right]\varphi_k\hspace{-0.5mm}\left[m,n\right]\right)$
		\STATE /* Expansion coefficient estimation */
		\STATE $\hat{c} = \gamma p_u$
		\STATE /* Model and residual update */
		\STATE $g\left[m,n\right] = g\left[m,n\right] + \hat{c}\varphi_u\left[m,n\right], \forall \left(m,n\right)$
		\STATE $r\left[m,n\right] = r\left[m,n\right] - \hat{c}\varphi_u\left[m,n\right], \forall \left(m,n\right)$
	\ENDFOR
	\STATE /* Replace distorted signal parts */ 
	\FORALL {$\left(m,n\right)\in\mathcal{B}$}
		\STATE $s\left[m,n\right] = g\left[m,n\right]$
	\ENDFOR
\OUTPUT extrapolated signal $s\left[m,n\right]$
\end{algorithmic}
\end{algorithm}


\section{Fast Selective Extrapolation} \label{sec:rse}

In order to solve the dilemma of the huge computational complexity of SE, we propose a novel formulation of this algorithm that also operates in the spatial domain but is as fast as transform domain algorithms which have been mentioned at the beginning. With that, the advantages of both approaches are combined: the high speed of transform domain algorithms and the independence from certain basis function sets, offered by the spatial domain SE algorithm. The high speed of the novel algorithm results from the fact that the weighted scalar products only have to be evaluated once, prior to the first iteration. In the successive iterations they can be replaced by a recursive calculation. The novel algorithm is called Fast Selective Extrapolation (FaSE) and is outlined in detail for the general complex-valued scenario subsequently. If only real-valued signals and basis functions are regarded, the conjugate complex operations can just be discarded.

Although the principal behavior of FaSE is similar to SE, not the residual $r\left[m,n\right]$ in the spatial domain is regarded, but rather the weighted scalar products between the residual and the basis functions. This yields
\begin{equation}
 \label{eq:scalar} R_k^{\left(\nu\right)} = \sum_{\left(m,n\right)\in\mathcal{L}} r^{\left(\nu\right)} \left[m,n\right] \varphi^\ast_k\left[m,n\right] w\left[m,n\right], \ \forall k
\end{equation}
for depicting the weighted scalar product between the residual and the basis function with index $k$ in the $\nu$-th iteration. This scalar product has to be evaluated only once explicitly. This has to be done for the initial step, where the residual is equal to the original signal, leading to
\begin{equation}
 \label{eq:scalar0} R_k^{\left(0\right)} = \sum_{\left(m,n\right)\in \mathcal{L}} s \left[m,n\right] \varphi_k^\ast\left[m,n\right] w\left[m,n\right], \ \forall k  \ .
\end{equation}
After the initial $R_k^{\left(0\right)}$ have been determined, all subsequent calculations can be carried out with respect to the weighted scalar products and no explicit evaluation of the scalar products is necessary anymore.

Using $R_k^{\left(\nu\right)}$ and exploiting the fact that the square root is a monotonic increasing function for positive arguments, the basis function selection from (\ref{eq:bf_selection_complex}) can be simplified to
\begin{equation}
 \label{eq:u_nu}u^{\left(\nu\right)} = \argmax_k \frac{ \left|R_k^{\left(\nu-1\right)}\right|}{\hspace{-1mm}\sqrt{\displaystyle\sum_{\left(m,n\right)\in\mathcal{L}}\hspace{-2mm} \varphi_k^\ast\left[m,n\right]w\left[m,n\right]\varphi_k\left[m,n\right]}}.
\end{equation}

Using expression $R_{u^{\left(\nu\right)}}^{\left(\nu-1\right)}$ for the weighted scalar product between the selected basis function and the residual from the previous iteration, the estimate for the expansion coefficient results to
\begin{equation}
\label{eq:ec_estimation} \hat{c}_{u^{\left(\nu\right)}} = \gamma \frac{R_{u^{\left(\nu\right)}}^{\left(\nu-1\right)}}{\displaystyle\sum_{\left(m,n\right)\in\mathcal{L}} \varphi_{u^{\left(\nu\right)}}^\ast\left[m,n\right]w\left[m,n\right]\varphi_{u^{\left(\nu\right)}}\left[m,n\right]}.
\end{equation}
Here, again fast orthogonality deficiency compensation is used to derive the estimate for the expansion coefficient from the projection coefficient. Finally, the update of the model in every iteration can be carried out according to (\ref{eq:model_update}).

For the subsequent iterations the weighted scalar products can be updated by applying definition (\ref{eq:scalar}) on the residual update from (\ref{eq:residual_update}), yielding
\begin{eqnarray}
\nonumber R_k^{\left(\nu\right)} \hspace{-4mm}&=&\hspace{-6mm} \sum_{\left(m,n\right)\in\mathcal{L}} \hspace{-4mm}\left(\hspace{-0.5mm}r^{\left(\nu-1\right)}\hspace{-0.5mm}\left[m,n\right] \hspace{-0.5mm}-\hspace{-0.5mm} \hat{c}_{u^{\left(\nu\right)}} \varphi_{u^{\left(\nu\right)}}\hspace{-0.5mm} \left[m,n\right]\hspace{-0.5mm}\right)\hspace{-0.5mm} \varphi^\ast_k\hspace{-0.5mm}\left[m,n\right] w\hspace{-0.5mm}\left[m,n\right]\\
\label{eq:xi_update} \hspace{-5mm}&=&\hspace{-3mm} R_k^{\left(\nu-1\right)} \hspace{-1mm}-\hspace{-1mm} \hat{c}_{u^{\left(\nu\right)}}\hspace{-4mm} \displaystyle\sum_{\left(m,n\right)\in\mathcal{L}} \hspace{-3mm} \varphi_k^\ast\left[m,n\right] w\left[m,n\right] \varphi_{u^{\left(\nu\right)}}\left[m,n\right].
\end{eqnarray}
Obviously, the weighted scalar product between the residual and a certain basis function can be easily updated from one iteration to the other by subtracting the weighted scalar product between the actual basis function and the selected one, further weighted by the estimated expansion coefficient. Since the update only incorporates the weighted scalar product between two basis functions and is independent of the actual residual, it can be carried out very fast by calculating the different weighted scalar products of all basis functions in advance.

This novel formulation of the SE algorithm has two advantages. First of all, the residual now does not have to be calculated explicitly in every iteration step anymore, rather the weighted scalar products between the residual and the basis functions are updated. But more important is the fact, that the most complex calculations can be carried out in advance and can be tabulated. Namely, these are the weighted scalar products between every two basis functions and one over the square root of the weighted scalar product between a basis function and itself. This leads to the matrix
\begin{equation}
\label{eq:table_generation}
 C_{\left(k,l \right)} = \displaystyle\sum_{\left(m,n\right)\in\mathcal{L}} \varphi_k^\ast\left[m,n\right]w\left[m,n\right]\varphi_l\left[m,n\right], \ \forall k,l
\end{equation}
containing the weighted scalar products between every two basis functions and the vector
\begin{equation}
 D_k \hspace{-1mm}=\hspace{-1mm} \frac{1}{\sqrt{\displaystyle\sum_{\left(m,n\right)\in\mathcal{L}}\hspace{-3mm} \varphi_k^\ast\left[m,n\right]w\left[m,n\right]\varphi_k\left[m,n\right]}}= \frac{1}{\sqrt{C_{\left(k,k\right)}}}, \ \forall k
\end{equation}
holding the inverse of the square root of the weighted scalar products. Obviously, $C_{\left(k,l \right)}$ and $D_k$ are independent of the input signal and the residual. Hence, they only have to be calculated once and do not have to be calculated for every extrapolation process. Thus, they can either be computed at the beginning of the extrapolation process or read from storage. During the whole computation, they are kept in memory. Furthermore, as $C_{\left(k,l \right)}$ is of size $\left|\mathfrak{D}\right|^2$ and $D_k$ has length $\left|\mathfrak{D}\right|$, the memory consumption is manageable without any problems. Here, the expression $\left|\mathfrak{D}\right|$ expresses the cardinality of dictionary $\mathfrak{D}$ that contains all possible basis functions. Regarding the two equations above, one can see that they both depend on the weighting function. If different weighting functions are used, $C_{\left(k,l \right)}$ and $D_k$ have to be adapted according to the weighting function. But, regarding typical signal extrapolation tasks as \mbox{e.\ g.\ } error concealment or prediction, the same patterns or only a small number of different patterns occur. Therefore, this also is no problem, as $C_{\left(k,l \right)}$ and $D_k$ can be calculated for the different patterns in advance as well. During the generation of $C_{\left(k,l \right)}$ the complex symmetry of this matrix can be exploited and only $\frac{\left|\mathfrak{D}\right|^2+\left|\mathfrak{D}\right|}{2}$ weighted scalar products have to be actually calculated.

Using these pre-calculated and tabulated values, the basis function selection from (\ref{eq:u_nu}) can be rewritten as
\begin{equation}
 u^{\left(\nu\right)} = \argmax_k { \left|R_k^{\left(\nu-1\right)}\right|}\cdot{D_k}.
\end{equation}
In addition to that, the estimation of the expansion coefficient from (\ref{eq:ec_estimation}) can also be expressed very compactly by
\begin{equation}
\hat{c}_{u^{\left(\nu\right)}} = \gamma {R_{u^{\left(\nu\right)}}^{\left(\nu-1\right)}}{D_{u^{\left(\nu\right)}}^2}.
\end{equation}
Furthermore, the update of the weighted scalar products between the residual and all possible basis functions from (\ref{eq:xi_update}) can also be formulated very efficiently by
\begin{equation}
 R_k^{\left(\nu\right)} = R_k^{\left(\nu-1\right)} - \hat{c}_{u^{\left(\nu\right)}} C_{\left(k,u^{\left(\nu\right)}\right)}, \ \forall k.
\end{equation}
Regarding the three equations above, one can recognize that instead of evaluating the weighted scalar products in every iteration step explicitly, only one value has to be read from memory for every calculation. Thus, the very high computational load from the original spatial domain SE is traded against an increased memory consumption. But as the memory consumption still is easily manageable this is a quite reasonable exchange. 

The novel FaSE implementation has the further advantage that no divisions are required. With that, this implementation is suited more for fixed point or integer implementations than the original SE. In such a scenario, $D_k$ could be calculated with high accuracy and then quantized to integer or fixed point values. Thus, no expensive divisions have to be carried out within the iteration loop and the effect of error propagation due to a restricted word length can be reduced. Depending on the architecture on which the extrapolation is carried out and the regarded application, it may be preferable to store $\frac{1}{C_{\left(k,k\right)}}$ instead of $\frac{1}{\sqrt{C_{\left(k,k\right)}}}$ and to calculate $\left|\cdot\right|^2$ instead of $\left|\cdot\right|$. By using this modification, the complexity could be reduced a little bit more, if the platform on which the extrapolation runs directly supports the relevant operations. Nevertheless, at this point a sufficiently high computational accuracy is assumed for the above outlined calculations. For a hardware implementation or an implementation on a digital signal processor, finite-word length effects have to be considered and further research is necessary for determining the required bit-depth of the tables and the impact of fixed-point arithmetic.

\begin{algorithm}[t]
\caption{Generation of the tabulated values $C_{\left(k,l \right)}$ and $D_k$}
\label{algo:table_generation}
\fontsize{9}{8}\selectfont
\begin{algorithmic}
\INPUT basis functions $\varphi_k\left[m,n\right]$, weighting function $w\left[m,n\right]$
	\FORALL {$k=0, \ldots, \left|\mathfrak{D}\right|-1$}
		\FORALL {$l=k, \ldots, \left|\mathfrak{D}\right|-1$}
		\STATE $C_{\left(k,l\right)} = \sum_{\left(m,n\right)\in\mathcal{L}} \varphi_k^\ast\left[m,n\right]w\left[m,n\right]\varphi_l\left[m,n\right]$
		\STATE $C_{\left(l,k\right)} = C_{\left(k,l\right)}^\ast$
		\ENDFOR
		\STATE $D_k = \frac{1}{\sqrt{C_{\left(k,k\right)}}}$
	\ENDFOR
\OUTPUT tabulated values $C_{\left(k,l \right)}$ and $D_k$
\end{algorithmic}
\end{algorithm}

\begin{algorithm}[t]
\caption{Fast Selective Extrapolation for arbitrary basis functions}
\label{algo:rse}
\fontsize{9}{8}\selectfont
\begin{algorithmic}
\INPUT distorted signal $s\left[m,n\right]$, weighting function $w\left[m,n\right]$, basis functions $\varphi_k\left[m,n\right]$, tabulated values $C_{\left(k,l \right)}$ and $D_k$
	\STATE /* Calculation of the initial weighted scalar product */
	\FORALL {$k=0, \ldots, \left|\mathfrak{D}\right|-1$}
		\STATE $R_k=\sum_{\left(m,n\right)\in\mathcal{L}} s\left[m,n\right] \varphi_k^\ast\left[m,n\right] w\left[m,n\right]$
	\ENDFOR
	\FORALL {$\nu=1, \ldots, I$}
		\STATE /* Basis function selection */
		\STATE $u = \argmax_k \left|R_k\right|D_k$
		\STATE /* Expansion coefficient estimation */
		\STATE $\hat{c} = \gamma R_u D_u^2$
		\STATE /* Model update */
		\STATE $g\left[m,n\right] = g\left[m,n\right] + \hat{c}\varphi_u\left[m,n\right], \forall \left(m,n\right)$
		\FORALL {$k=0,\ldots,\left|\mathfrak{D}\right|-1$}
			\STATE $R_k = R_k - \hat{c} C_{\left(k,u\right)}$
		\ENDFOR
	\ENDFOR
	\STATE /* Replace distorted signal parts */ 
	\FORALL {$\left(m,n\right)\in\mathcal{B}$}
		\STATE $s\left[m,n\right] = g\left[m,n\right]$
	\ENDFOR
\OUTPUT extrapolated signal $s\left[m,n\right]$
\end{algorithmic}
\end{algorithm}

In order to give a final overview of FaSE, Alg.\ \ref{algo:table_generation} and \ref{algo:rse} show the pseudo code for generating the tabulated values and for the actual model generation. The table generation is separated from the model generation for emphasizing again that the generation of the tables only has to be carried out once. Regarding the operations that have to be carried out within the iteration loop, one can recognize that only very simple operations have to be performed which can furthermore be processed very fast. The only computational expensive operation is the initial calculation of $R_k^{\left(0\right)}$, but compared to the original SE, this complex step only has to be carried out only once instead of in every iteration.


\section{Complexity Evaluation}\label{sec:complexity} 

Regarding the two previous sections, one can recognize that FaSE is able to outperform the original SE since the computational complexity within the iteration loop is reduced and since as many calculations as possible are carried out in advance and are tabulated. To quantify the complexity of SE and FaSE, the number of operations is regarded that is necessary for generating the model by each of the algorithms. In Tab.\ \ref{tab:complexity} for SE, FaSE and the table generation for FaSE, the number of operations is listed, depending on the extent $M,N$ of extrapolation area $\mathcal{L}$, dictionary size $\left|\mathfrak{D}\right|$ and the number of iterations $I$ to be carried out. Here, the operations are separated into three groups, the number of multiplications (MUL), the number of additions (ADD) and the number of other operations (OTHER) like divisions, comparisons or the calculation of a square root. As the general case of complex-valued signals and basis functions is regarded, MUL and ADD describe complex-valued multiplications and additions. For presentational reasons, a further separation of these operations into real-valued operations is omitted. 

\begin{table}
\centering
\caption{Number of required operations for model generation by SE and FaSE and for generating the tables.}
\begin{tabular}{|c|c|}
\hline
\multicolumn{2}{|c|}{SE}  \\ \hline
MUL & $I\cdot\left(6MN\cdot\left|\mathfrak{D}\right| + \left|\mathfrak{D}\right|+2MN+1\right)$ \\ \hline
ADD & $I\cdot\left(3MN\cdot\left|\mathfrak{D}\right| + 2MN\right)$ \\ \hline
OTHER & $3I\cdot\left|\mathfrak{D}\right|$ \\ \hline \multicolumn{2}{c}{ }\\ \hline  
\multicolumn{2}{|c|}{FaSE}  \\ \hline
MUL & $2MN\cdot\left|\mathfrak{D}\right| + I\cdot\left(2\left|\mathfrak{D}\right|+MN+1\right)$ \\ \hline
ADD & $MN\cdot\left|\mathfrak{D}\right| + I\cdot\left(\left|\mathfrak{D}\right|+MN\right)$ \\ \hline
OTHER & $2I\cdot\left|\mathfrak{D}\right|$ \\ \hline\hline
\multicolumn{2}{|c|}{FaSE Table Generation}  \\ \hline
MUL & $\left(\left|\mathfrak{D}\right|^2+\left|\mathfrak{D}\right|\right) \cdot MN$ \\ \hline
ADD & $\left(\left|\mathfrak{D}\right|^2+\left|\mathfrak{D}\right|\right) \cdot MN / 2$ \\ \hline
OTHER & $\left(\left|\mathfrak{D}\right|^2+\left|\mathfrak{D}\right|\right) \cdot MN / 2 + \left|\mathfrak{D}\right|$ \\ \hline
\end{tabular}
\label{tab:complexity}
\end{table}

\begin{figure}
	\centering



\providelength{\AxesLineWidth}       \setlength{\AxesLineWidth}{0.5pt}
\providelength{\GridLineWidth}       \setlength{\GridLineWidth}{0.4pt}
\providelength{\GridLineDotSep}      \setlength{\GridLineDotSep}{0.4pt}
\providelength{\MinorGridLineWidth}  \setlength{\MinorGridLineWidth}{0.4pt}
\providelength{\MinorGridLineDotSep} \setlength{\MinorGridLineDotSep}{0.8pt}
\providelength{\plotwidth}           \setlength{\plotwidth}{7cm} 
\providelength{\LineWidth}           \setlength{\LineWidth}{0.7pt}
\providelength{\MarkerSize}          \setlength{\MarkerSize}{6pt}
\newrgbcolor{GridColor}{0.8 0.8 0.8}

\psset{xunit=0.000250\plotwidth,yunit=0.1153333\plotwidth}
\begin{pspicture}(-571.428571,5.196429)(4000.000000,12.000000)


\psline[linestyle=dashed,dash=2pt 3pt,dotsep=\GridLineDotSep,linewidth=\GridLineWidth,linecolor=GridColor](0.000000,6.000000)(0.000000,12.000000)
\psline[linestyle=dashed,dash=2pt 3pt,dotsep=\GridLineDotSep,linewidth=\GridLineWidth,linecolor=GridColor](500.000000,6.000000)(500.000000,12.000000)
\psline[linestyle=dashed,dash=2pt 3pt,dotsep=\GridLineDotSep,linewidth=\GridLineWidth,linecolor=GridColor](1000.000000,6.000000)(1000.000000,12.000000)
\psline[linestyle=dashed,dash=2pt 3pt,dotsep=\GridLineDotSep,linewidth=\GridLineWidth,linecolor=GridColor](1500.000000,6.000000)(1500.000000,12.000000)
\psline[linestyle=dashed,dash=2pt 3pt,dotsep=\GridLineDotSep,linewidth=\GridLineWidth,linecolor=GridColor](2000.000000,6.000000)(2000.000000,12.000000)
\psline[linestyle=dashed,dash=2pt 3pt,dotsep=\GridLineDotSep,linewidth=\GridLineWidth,linecolor=GridColor](2500.000000,6.000000)(2500.000000,12.000000)
\psline[linestyle=dashed,dash=2pt 3pt,dotsep=\GridLineDotSep,linewidth=\GridLineWidth,linecolor=GridColor](3000.000000,6.000000)(3000.000000,12.000000)
\psline[linestyle=dashed,dash=2pt 3pt,dotsep=\GridLineDotSep,linewidth=\GridLineWidth,linecolor=GridColor](3500.000000,6.000000)(3500.000000,12.000000)
\psline[linestyle=dashed,dash=2pt 3pt,dotsep=\GridLineDotSep,linewidth=\GridLineWidth,linecolor=GridColor](4000.000000,6.000000)(4000.000000,12.000000)
\psline[linestyle=dashed,dash=2pt 3pt,dotsep=\GridLineDotSep,linewidth=\GridLineWidth,linecolor=GridColor](0.000000,6.000000)(4000.000000,6.000000)
\psline[linestyle=dashed,dash=2pt 3pt,dotsep=\GridLineDotSep,linewidth=\GridLineWidth,linecolor=GridColor](0.000000,7.000000)(4000.000000,7.000000)
\psline[linestyle=dashed,dash=2pt 3pt,dotsep=\GridLineDotSep,linewidth=\GridLineWidth,linecolor=GridColor](0.000000,8.000000)(4000.000000,8.000000)
\psline[linestyle=dashed,dash=2pt 3pt,dotsep=\GridLineDotSep,linewidth=\GridLineWidth,linecolor=GridColor](0.000000,9.000000)(4000.000000,9.000000)
\psline[linestyle=dashed,dash=2pt 3pt,dotsep=\GridLineDotSep,linewidth=\GridLineWidth,linecolor=GridColor](0.000000,10.000000)(4000.000000,10.000000)
\psline[linestyle=dashed,dash=2pt 3pt,dotsep=\GridLineDotSep,linewidth=\GridLineWidth,linecolor=GridColor](0.000000,11.000000)(4000.000000,11.000000)
\psline[linestyle=dashed,dash=2pt 3pt,dotsep=\GridLineDotSep,linewidth=\GridLineWidth,linecolor=GridColor](0.000000,12.000000)(4000.000000,12.000000)

\psline[linewidth=\AxesLineWidth,linecolor=GridColor](0.000000,6.000000)(0.000000,6.090000)
\psline[linewidth=\AxesLineWidth,linecolor=GridColor](500.000000,6.000000)(500.000000,6.090000)
\psline[linewidth=\AxesLineWidth,linecolor=GridColor](1000.000000,6.000000)(1000.000000,6.090000)
\psline[linewidth=\AxesLineWidth,linecolor=GridColor](1500.000000,6.000000)(1500.000000,6.090000)
\psline[linewidth=\AxesLineWidth,linecolor=GridColor](2000.000000,6.000000)(2000.000000,6.090000)
\psline[linewidth=\AxesLineWidth,linecolor=GridColor](2500.000000,6.000000)(2500.000000,6.090000)
\psline[linewidth=\AxesLineWidth,linecolor=GridColor](3000.000000,6.000000)(3000.000000,6.090000)
\psline[linewidth=\AxesLineWidth,linecolor=GridColor](3500.000000,6.000000)(3500.000000,6.090000)
\psline[linewidth=\AxesLineWidth,linecolor=GridColor](4000.000000,6.000000)(4000.000000,6.090000)
\psline[linewidth=\AxesLineWidth,linecolor=GridColor](0.000000,6.000000)(48.000000,6.000000)
\psline[linewidth=\AxesLineWidth,linecolor=GridColor](0.000000,7.000000)(48.000000,7.000000)
\psline[linewidth=\AxesLineWidth,linecolor=GridColor](0.000000,8.000000)(48.000000,8.000000)
\psline[linewidth=\AxesLineWidth,linecolor=GridColor](0.000000,9.000000)(48.000000,9.000000)
\psline[linewidth=\AxesLineWidth,linecolor=GridColor](0.000000,10.000000)(48.000000,10.000000)
\psline[linewidth=\AxesLineWidth,linecolor=GridColor](0.000000,11.000000)(48.000000,11.000000)
\psline[linewidth=\AxesLineWidth,linecolor=GridColor](0.000000,12.000000)(48.000000,12.000000)

{ \footnotesize 
\rput[t](0.000000,5.910000){$0$}
\rput[t](500.000000,5.910000){$500$}
\rput[t](1000.000000,5.910000){$1000$}
\rput[t](1500.000000,5.910000){$1500$}
\rput[t](2000.000000,5.910000){$2000$}
\rput[t](2500.000000,5.910000){$2500$}
\rput[t](3000.000000,5.910000){$3000$}
\rput[t](3500.000000,5.910000){$3500$}
\rput[t](4000.000000,5.910000){$4000$}
\rput[r](-48.000000,6.000000){$10^{6}$}
\rput[r](-48.000000,7.000000){$10^{7}$}
\rput[r](-48.000000,8.000000){$10^{8}$}
\rput[r](-48.000000,9.000000){$10^{9}$}
\rput[r](-48.000000,10.000000){$10^{10}$}
\rput[r](-48.000000,11.000000){$10^{11}$}
\rput[r](-48.000000,12.000000){$10^{12}$}
} 

\pspolygon[linewidth=\AxesLineWidth](0.000000,6.000000)(4000.000000,6.000000)(4000.000000,12.000000)(0.000000,12.000000)(0.000000,6.000000)

{ \small 
\rput[b](2000.000000,5.196429){
\begin{tabular}{c}
Iterations\\
\end{tabular}
}

\rput[t]{90}(-571.428571,9.000000){
\begin{tabular}{c}
Operations\\
\end{tabular}
}
} 

\savedata{\mydata}[{
{1.000000,8.179057},{10.000000,9.179057},{50.000000,9.878027},{150.000000,10.355148},{250.000000,10.576997},
{400.000000,10.781117},{750.000000,11.054118},{1000.000000,11.179057},{1500.000000,11.355148},{2000.000000,11.480087},
{3000.000000,11.656178},{4000.000000,11.781117}
}]

\newrgbcolor{color233.0183}{0  0  1}
\savedata{\mydata}[{
{1.000000,8.179057},{10.000000,9.179057},{50.000000,9.878027},{150.000000,10.355148},{250.000000,10.576997},
{400.000000,10.781117},{750.000000,11.054118},{1000.000000,11.179057},{1500.000000,11.355148},{2000.000000,11.480087},
{3000.000000,11.656178},{4000.000000,11.781117}
}]
\dataplot[plotstyle=line,showpoints=true,dotstyle=asterisk,dotsize=\MarkerSize,linestyle=solid,linewidth=\LineWidth,linecolor=color233.0183]{\mydata}

\newrgbcolor{color234.0168}{0.1  1  0.7}
\savedata{\mydata}[{
{1.000000,11.138216},{10.000000,11.138216},{50.000000,11.138216},{150.000000,11.138216},{300.000000,11.138216},
{750.000000,11.138216},{1000.000000,11.138216},{1500.000000,11.138216},{2000.000000,11.138216},{3000.000000,11.138216},
{4000.000000,11.138216}
}]
\dataplot[plotstyle=line,linestyle=dashed,linewidth=\LineWidth,linecolor=color234.0168]{\mydata}

\savedata{\mydata}[{
{1.000000,11.138216}{1.000000,11.138216}
}]
\dataplot[plotstyle=line,showpoints=true,dotstyle=diamond*,dotsize=8pt,linestyle=dashed,linewidth=\LineWidth,linecolor=color234.0168]{\mydata}

\newrgbcolor{color235.0168}{1  0  0}
\savedata{\mydata}[{
{1.000000,7.702088},{10.000000,7.704308},{50.000000,7.714039},{150.000000,7.737452},{300.000000,7.770363},
{750.000000,7.856345},{1000.000000,7.897653},{1500.000000,7.970073},{2000.000000,8.032126},{3000.000000,8.134657},
{4000.000000,8.217546}
}]
\dataplot[plotstyle=line,showpoints=true,dotstyle=o,dotsize=\MarkerSize,linestyle=solid,linewidth=\LineWidth,linecolor=color235.0168]{\mydata}

{ \small 
\rput[br](3904.000000,6.180000){%
\psshadowbox[framesep=0pt,linewidth=\AxesLineWidth]{\psframebox*{\begin{tabular}{l}
\Rnode{a1}{\hspace*{0.0ex}} \hspace*{0.7cm} \Rnode{a2}{~~SE} \\
\Rnode{a5}{\hspace*{0.0ex}} \hspace*{0.7cm} \Rnode{a6}{~~FaSE} \\
\Rnode{a3}{\hspace*{0.0ex}} \hspace*{0.7cm} \Rnode{a4}{~~Table Generation} \\
\end{tabular}}
\ncline[linestyle=solid,linewidth=\LineWidth,linecolor=color233.0183]{a1}{a2} \ncput{\psdot[dotstyle=asterisk,dotsize=\MarkerSize,linecolor=color233.0183]}
\ncline[linestyle=dashed,linewidth=\LineWidth,linecolor=color234.0168]{a3}{a4}\ncput{\psdot[dotstyle=diamond*,dotsize=8pt,linecolor=color234.0168]}
\ncline[linestyle=solid,linewidth=\LineWidth,linecolor=color235.0168]{a5}{a6} \ncput{\psdot[dotstyle=o,dotsize=\MarkerSize,linecolor=color235.0168]}
}%
}%
} 

\end{pspicture}%
\vspace{-2mm}
	\caption{Operations per block for model generation by SE and FaSE and operations necessary for generating tabulated $C_{\left(k,l\right)}$ and $D_k$.  For comparison, the operations for generating the tables are drawn over the complete iterations range although they have to be calculated only once. Spatial sizes $M=64$ and $N=64$ and dictionary size $\left|\mathfrak{D}\right|=4096$.\vspace{-3mm}}
	\label{fig:complexity}
\end{figure}
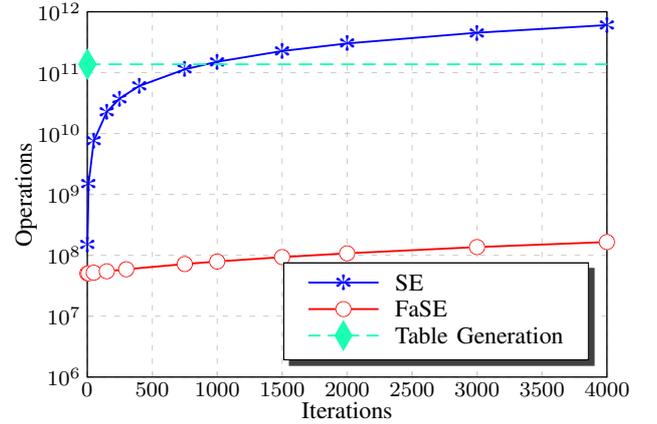

The computationally most expensive step in SE is the projection onto the basis functions. For the weighted projection of the residual onto a single basis function, $4MN$ complex-valued multiplications, $2MN$ additions and one division is required. Since SE has to project the residual in every iteration onto every of the $\left|\mathfrak{D}\right|$ basis functions, these numbers have to be further multiplied by $I\cdot\left|\mathfrak{D}\right|$. For selecting the basis function to be added, in every iteration $\left(2MN+1\right)\left|\mathfrak{D}\right|$ multiplications, $MN\left|\mathfrak{D}\right|$ additions and $\left|\mathfrak{D}\right|$ comparisons and absolute value calculations are required and the model and residual update consumes $2MN+1$ multiplications and $2MN$ additions. Due to this, the overall complexity of SE with respect to the number of iterations is proportional to  $\mathcal{O}\left(I\cdot MN\cdot \left|\mathfrak{D}\right|\right)$. In contrast to this, FaSE has to evaluate the weighted scalar product between the input signal and the basis functions only once, prior to the iterations. This calculation requires only $2MN\cdot\left|\mathfrak{D}\right|$ complex-valued multiplications and $MN\cdot\left|\mathfrak{D}\right|$ additions. Within every iteration, only 
$2\left|\mathfrak{D}\right|+MN+1$ complex-valued multiplications, $\left|\mathfrak{D}\right|+MN$ additions, and $\left|\mathfrak{D}\right|$ comparisons and absolute value calculations have to be carried out. As $\left|\mathfrak{D}\right|$ and $MN$ are of the same magnitude, the computational complexity of FaSE increases proportional to $\mathcal{O}\left( I \cdot \left|\mathfrak{D}\right|\right)$ with respect to the number of iterations. For generating the tables, one has to consider, that the weighted scalar products between every two basis functions have to be evaluated, resulting in a complexity that is proportional to  $\mathcal{O}\left(MN\cdot \left|\mathfrak{D}\right|^2\right)$, as shown in Tab.\ \ref{tab:complexity}.

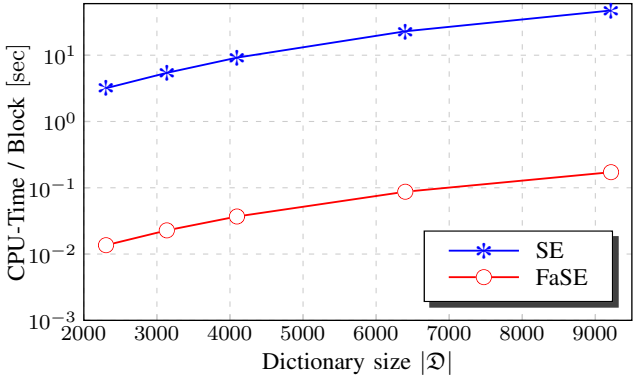
\begin{figure}
	\centering



\providelength{\AxesLineWidth}       \setlength{\AxesLineWidth}{0.5pt}
\providelength{\GridLineWidth}       \setlength{\GridLineWidth}{0.4pt}
\providelength{\GridLineDotSep}      \setlength{\GridLineDotSep}{0.4pt}
\providelength{\MinorGridLineWidth}  \setlength{\MinorGridLineWidth}{0.4pt}
\providelength{\MinorGridLineDotSep} \setlength{\MinorGridLineDotSep}{0.8pt}
\providelength{\plotwidth}           \setlength{\plotwidth}{7cm} 
\providelength{\LineWidth}           \setlength{\LineWidth}{0.7pt}
\providelength{\MarkerSize}          \setlength{\MarkerSize}{6pt}
\newrgbcolor{GridColor}{0.8 0.8 0.8}

\psset{xunit=0.000133\plotwidth,yunit=0.125572\plotwidth}
\begin{pspicture}(928.571429,-3.853241)(9500.000000,1.778151)


\psline[linestyle=dashed,dash=2pt 3pt,dotsep=\GridLineDotSep,linewidth=\GridLineWidth,linecolor=GridColor](2000.000000,-3.000000)(2000.000000,1.778151)
\psline[linestyle=dashed,dash=2pt 3pt,dotsep=\GridLineDotSep,linewidth=\GridLineWidth,linecolor=GridColor](3000.000000,-3.000000)(3000.000000,1.778151)
\psline[linestyle=dashed,dash=2pt 3pt,dotsep=\GridLineDotSep,linewidth=\GridLineWidth,linecolor=GridColor](4000.000000,-3.000000)(4000.000000,1.778151)
\psline[linestyle=dashed,dash=2pt 3pt,dotsep=\GridLineDotSep,linewidth=\GridLineWidth,linecolor=GridColor](5000.000000,-3.000000)(5000.000000,1.778151)
\psline[linestyle=dashed,dash=2pt 3pt,dotsep=\GridLineDotSep,linewidth=\GridLineWidth,linecolor=GridColor](6000.000000,-3.000000)(6000.000000,1.778151)
\psline[linestyle=dashed,dash=2pt 3pt,dotsep=\GridLineDotSep,linewidth=\GridLineWidth,linecolor=GridColor](7000.000000,-3.000000)(7000.000000,1.778151)
\psline[linestyle=dashed,dash=2pt 3pt,dotsep=\GridLineDotSep,linewidth=\GridLineWidth,linecolor=GridColor](8000.000000,-3.000000)(8000.000000,1.778151)
\psline[linestyle=dashed,dash=2pt 3pt,dotsep=\GridLineDotSep,linewidth=\GridLineWidth,linecolor=GridColor](9000.000000,-3.000000)(9000.000000,1.778151)
\psline[linestyle=dashed,dash=2pt 3pt,dotsep=\GridLineDotSep,linewidth=\GridLineWidth,linecolor=GridColor](2000.000000,-3.000000)(9500.000000,-3.000000)
\psline[linestyle=dashed,dash=2pt 3pt,dotsep=\GridLineDotSep,linewidth=\GridLineWidth,linecolor=GridColor](2000.000000,-2.000000)(9500.000000,-2.000000)
\psline[linestyle=dashed,dash=2pt 3pt,dotsep=\GridLineDotSep,linewidth=\GridLineWidth,linecolor=GridColor](2000.000000,-1.000000)(9500.000000,-1.000000)
\psline[linestyle=dashed,dash=2pt 3pt,dotsep=\GridLineDotSep,linewidth=\GridLineWidth,linecolor=GridColor](2000.000000,0.000000)(9500.000000,0.000000)
\psline[linestyle=dashed,dash=2pt 3pt,dotsep=\GridLineDotSep,linewidth=\GridLineWidth,linecolor=GridColor](2000.000000,1.000000)(9500.000000,1.000000)

\psline[linewidth=\AxesLineWidth,linecolor=GridColor](2000.000000,-3.000000)(2000.000000,-2.904437)
\psline[linewidth=\AxesLineWidth,linecolor=GridColor](3000.000000,-3.000000)(3000.000000,-2.904437)
\psline[linewidth=\AxesLineWidth,linecolor=GridColor](4000.000000,-3.000000)(4000.000000,-2.904437)
\psline[linewidth=\AxesLineWidth,linecolor=GridColor](5000.000000,-3.000000)(5000.000000,-2.904437)
\psline[linewidth=\AxesLineWidth,linecolor=GridColor](6000.000000,-3.000000)(6000.000000,-2.904437)
\psline[linewidth=\AxesLineWidth,linecolor=GridColor](7000.000000,-3.000000)(7000.000000,-2.904437)
\psline[linewidth=\AxesLineWidth,linecolor=GridColor](8000.000000,-3.000000)(8000.000000,-2.904437)
\psline[linewidth=\AxesLineWidth,linecolor=GridColor](9000.000000,-3.000000)(9000.000000,-2.904437)
\psline[linewidth=\AxesLineWidth,linecolor=GridColor](2000.000000,-3.000000)(2090.000000,-3.000000)
\psline[linewidth=\AxesLineWidth,linecolor=GridColor](2000.000000,-2.000000)(2090.000000,-2.000000)
\psline[linewidth=\AxesLineWidth,linecolor=GridColor](2000.000000,-1.000000)(2090.000000,-1.000000)
\psline[linewidth=\AxesLineWidth,linecolor=GridColor](2000.000000,0.000000)(2090.000000,0.000000)
\psline[linewidth=\AxesLineWidth,linecolor=GridColor](2000.000000,1.000000)(2090.000000,1.000000)

{ \footnotesize 
\rput[t](2000.000000,-3.095563){$2000$}
\rput[t](3000.000000,-3.095563){$3000$}
\rput[t](4000.000000,-3.095563){$4000$}
\rput[t](5000.000000,-3.095563){$5000$}
\rput[t](6000.000000,-3.095563){$6000$}
\rput[t](7000.000000,-3.095563){$7000$}
\rput[t](8000.000000,-3.095563){$8000$}
\rput[t](9000.000000,-3.095563){$9000$}
\rput[r](1910.000000,-3.000000){$10^{-3}$}
\rput[r](1910.000000,-2.000000){$10^{-2}$}
\rput[r](1910.000000,-1.000000){$10^{-1}$}
\rput[r](1910.000000,0.000000){$10^{0}$}
\rput[r](1910.000000,1.000000){$10^{1}$}
} 

\pspolygon[linewidth=\AxesLineWidth](2000.000000,-3.000000)(9500.000000,-3.000000)(9500.000000,1.778151)(2000.000000,1.778151)(2000.000000,-3.000000)

{ \small 
\rput[b](5750.000000,-3.853241){
\begin{tabular}{c}
Dictionary size $\left|\mathfrak{D}\right|$\\
\end{tabular}
}

\rput[t]{90}(928.571429,-0.610924){
\begin{tabular}{c}
CPU-Time / Block $\left[\mathrm{sec}\right]$\\
\end{tabular}
}
} 

\newrgbcolor{color457.0134}{1  0  0}
\savedata{\mydata}[{
{2304.000000,-1.866562},{3136.000000,-1.642730},{4096.000000,-1.433376},{6400.000000,-1.059863},{9216.000000,-0.764954},
}]
\dataplot[plotstyle=line,showpoints=true,dotstyle=o,dotsize=\MarkerSize,linestyle=solid,linewidth=\LineWidth,linecolor=color457.0134]{\mydata}

\newrgbcolor{color458.0129}{0  0  1}
\savedata{\mydata}[{
{2304.000000,0.501963},{3136.000000,0.733735},{4096.000000,0.963304},{6400.000000,1.360713},{9216.000000,1.674555},
}]
\dataplot[plotstyle=line,showpoints=true,dotstyle=asterisk,dotsize=\MarkerSize,linestyle=solid,linewidth=\LineWidth,linecolor=color458.0129]{\mydata}

{ \small 
\rput[br](9320.000000,-2.808874){%
\psshadowbox[framesep=0pt,linewidth=\AxesLineWidth]{\psframebox*{\begin{tabular}{l}
\Rnode{a3}{\hspace*{0.0ex}} \hspace*{0.7cm} \Rnode{a4}{~~SE} \\
\Rnode{a1}{\hspace*{0.0ex}} \hspace*{0.7cm} \Rnode{a2}{~~FaSE} \\
\end{tabular}}
\ncline[linestyle=solid,linewidth=\LineWidth,linecolor=color457.0134]{a1}{a2} \ncput{\psdot[dotstyle=o,dotsize=\MarkerSize,linecolor=color457.0134]}
\ncline[linestyle=solid,linewidth=\LineWidth,linecolor=color458.0129]{a3}{a4} \ncput{\psdot[dotstyle=asterisk,dotsize=\MarkerSize,linecolor=color458.0129]}
}%
}%
} 

\end{pspicture}%
\vspace{-2mm}
	\caption{Processing time over dictionary size for 2D model generation with arbitrary real-valued basis functions and $250$ iterations. The size of the extrapolation area is chosen so that $MN = \left|\mathfrak{D}\right|$ holds for every data point.\vspace{-3mm}}
	\label{fig:dct_over_N}
\end{figure}

Fig.\ \ref{fig:complexity} shows the number of operations with respect to the number of iterations for $M=N=64$ and $\left|\mathcal{D}\right|=4096$. This plot only shows the overall number of operations, i.\ e.\ the sum of MUL, ADD and OTHER, in order to give a rough impression of the overall complexity and compare the different algorithms. The fact that complex operations like divisions require more processing time than a simple multiplication is omitted for this plot. It can be easily recognized that the number of operations that is necessary for generating the model by SE is several decades larger than for FaSE. The plot further shows the number of operations that is required for generating the tabulated $C_{\left(k,l\right)}$ and $D_k$, indicated by a rhomb. In addition to that, the number of operations for the table generation is displayed as dashed line over the complete iteration range. It has to be noted that the table generation is independent of the iterations and this illustration is only chosen for comparing the complexity of the table generation with SE. Therewith, it can be recognized that the table generation requires roughly as many operations as $1000$ iterations of SE would require. Since the number of iterations for generating the model can easily reach values larger than $200$ as has been shown in \cite{Seiler2008}, the expense for generating the tables amortize even after a small number of blocks. Taking into account that in typical scenarios a large number of blocks is extrapolated with the same weighting function, the complexity for generating the tables very soon becomes negligible.


\section{Results for Arbitrary Basis Functions}\label{sec:results_arbitrary} 

\begin{figure}
	\centering



\providelength{\AxesLineWidth}       \setlength{\AxesLineWidth}{0.5pt}
\providelength{\GridLineWidth}       \setlength{\GridLineWidth}{0.4pt}
\providelength{\GridLineDotSep}      \setlength{\GridLineDotSep}{0.4pt}
\providelength{\MinorGridLineWidth}  \setlength{\MinorGridLineWidth}{0.4pt}
\providelength{\MinorGridLineDotSep} \setlength{\MinorGridLineDotSep}{0.8pt}
\providelength{\plotwidth}           \setlength{\plotwidth}{7cm} 
\providelength{\LineWidth}           \setlength{\LineWidth}{0.7pt}
\providelength{\MarkerSize}          \setlength{\MarkerSize}{6pt}
\newrgbcolor{GridColor}{0.8 0.8 0.8}

\psset{xunit=0.004000\plotwidth,yunit=0.162207\plotwidth}
\begin{pspicture}(-35.714286,-3.359500)(250.000000,1.000000)


\psline[linestyle=dashed,dash=2pt 3pt,dotsep=\GridLineDotSep,linewidth=\GridLineWidth,linecolor=GridColor](0.000000,-2.698970)(0.000000,1.000000)
\psline[linestyle=dashed,dash=2pt 3pt,dotsep=\GridLineDotSep,linewidth=\GridLineWidth,linecolor=GridColor](50.000000,-2.698970)(50.000000,1.000000)
\psline[linestyle=dashed,dash=2pt 3pt,dotsep=\GridLineDotSep,linewidth=\GridLineWidth,linecolor=GridColor](100.000000,-2.698970)(100.000000,1.000000)
\psline[linestyle=dashed,dash=2pt 3pt,dotsep=\GridLineDotSep,linewidth=\GridLineWidth,linecolor=GridColor](150.000000,-2.698970)(150.000000,1.000000)
\psline[linestyle=dashed,dash=2pt 3pt,dotsep=\GridLineDotSep,linewidth=\GridLineWidth,linecolor=GridColor](200.000000,-2.698970)(200.000000,1.000000)
\psline[linestyle=dashed,dash=2pt 3pt,dotsep=\GridLineDotSep,linewidth=\GridLineWidth,linecolor=GridColor](250.000000,-2.698970)(250.000000,1.000000)
\psline[linestyle=dashed,dash=2pt 3pt,dotsep=\GridLineDotSep,linewidth=\GridLineWidth,linecolor=GridColor](0.000000,-2.000000)(250.000000,-2.000000)
\psline[linestyle=dashed,dash=2pt 3pt,dotsep=\GridLineDotSep,linewidth=\GridLineWidth,linecolor=GridColor](0.000000,-1.000000)(250.000000,-1.000000)
\psline[linestyle=dashed,dash=2pt 3pt,dotsep=\GridLineDotSep,linewidth=\GridLineWidth,linecolor=GridColor](0.000000,0.000000)(250.000000,0.000000)
\psline[linestyle=dashed,dash=2pt 3pt,dotsep=\GridLineDotSep,linewidth=\GridLineWidth,linecolor=GridColor](0.000000,1.000000)(250.000000,1.000000)

\psline[linewidth=\AxesLineWidth,linecolor=GridColor](0.000000,-2.698970)(0.000000,-2.624991)
\psline[linewidth=\AxesLineWidth,linecolor=GridColor](50.000000,-2.698970)(50.000000,-2.624991)
\psline[linewidth=\AxesLineWidth,linecolor=GridColor](100.000000,-2.698970)(100.000000,-2.624991)
\psline[linewidth=\AxesLineWidth,linecolor=GridColor](150.000000,-2.698970)(150.000000,-2.624991)
\psline[linewidth=\AxesLineWidth,linecolor=GridColor](200.000000,-2.698970)(200.000000,-2.624991)
\psline[linewidth=\AxesLineWidth,linecolor=GridColor](250.000000,-2.698970)(250.000000,-2.624991)
\psline[linewidth=\AxesLineWidth,linecolor=GridColor](0.000000,-2.000000)(3.000000,-2.000000)
\psline[linewidth=\AxesLineWidth,linecolor=GridColor](0.000000,-1.000000)(3.000000,-1.000000)
\psline[linewidth=\AxesLineWidth,linecolor=GridColor](0.000000,0.000000)(3.000000,0.000000)
\psline[linewidth=\AxesLineWidth,linecolor=GridColor](0.000000,1.000000)(3.000000,1.000000)

{ \footnotesize 
\rput[t](0.000000,-2.772949){$0$}
\rput[t](50.000000,-2.772949){$50$}
\rput[t](100.000000,-2.772949){$100$}
\rput[t](150.000000,-2.772949){$150$}
\rput[t](200.000000,-2.772949){$200$}
\rput[t](250.000000,-2.772949){$250$}
\rput[r](-3.000000,-2.000000){$10^{-2}$}
\rput[r](-3.000000,-1.000000){$10^{-1}$}
\rput[r](-3.000000,0.000000){$10^{0}$}
\rput[r](-3.000000,1.000000){$10^{1}$}
} 

\pspolygon[linewidth=\AxesLineWidth](0.000000,-2.698970)(250.000000,-2.698970)(250.000000,1.000000)(0.000000,1.000000)(0.000000,-2.698970)

{ \small 
\rput[b](125.000000,-3.359500){
\begin{tabular}{c}
Iterations\\
\end{tabular}
}

\rput[t]{90}(-35.714286,-0.849485){
\begin{tabular}{c}
CPU-Time / Block $\left[\mathrm{sec}\right]$\\
\end{tabular}
}
} 

\newrgbcolor{color229.0134}{1  0  0}
\savedata{\mydata}[{
{5.000000,-1.521785},{10.000000,-1.521869},{25.000000,-1.511975},{50.000000,-1.504601},{75.000000,-1.495191},
{100.000000,-1.483726},{150.000000,-1.468504},{250.000000,-1.433376}
}]
\dataplot[plotstyle=line,showpoints=true,dotstyle=o,dotsize=\MarkerSize,linestyle=solid,linewidth=\LineWidth,linecolor=color229.0134]{\mydata}

\newrgbcolor{color230.0129}{0  0  1}
\savedata{\mydata}[{
{5.000000,-0.733645},{10.000000,-0.434562},{25.000000,-0.037153},{50.000000,0.263905},{75.000000,0.440430},
{100.000000,0.566455},{150.000000,0.741175},{250.000000,0.963304}
}]
\dataplot[plotstyle=line,showpoints=true,dotstyle=asterisk,dotsize=\MarkerSize,linestyle=solid,linewidth=\LineWidth,linecolor=color230.0129]{\mydata}

{ \small 
\rput[br](244.000000,-2.551011){%
\psshadowbox[framesep=0pt,linewidth=\AxesLineWidth]{\psframebox*{\begin{tabular}{l}
\Rnode{a3}{\hspace*{0.0ex}} \hspace*{0.7cm} \Rnode{a4}{~~SE} \\
\Rnode{a1}{\hspace*{0.0ex}} \hspace*{0.7cm} \Rnode{a2}{~~FaSE} \\
\end{tabular}}
\ncline[linestyle=solid,linewidth=\LineWidth,linecolor=color229.0134]{a1}{a2} \ncput{\psdot[dotstyle=o,dotsize=\MarkerSize,linecolor=color229.0134]}
\ncline[linestyle=solid,linewidth=\LineWidth,linecolor=color230.0129]{a3}{a4} \ncput{\psdot[dotstyle=asterisk,dotsize=\MarkerSize,linecolor=color230.0129]}
}%
}%
} 

\end{pspicture}%
\vspace{-2mm}
	\caption{Processing time over iterations for 2D model generation with arbitrary real-valued basis functions of size $64\times 64$ and dictionary size $\left|\mathfrak{D}\right|=4096$.\vspace{-1mm}}
	\label{fig:dct_over_iter}
\end{figure}
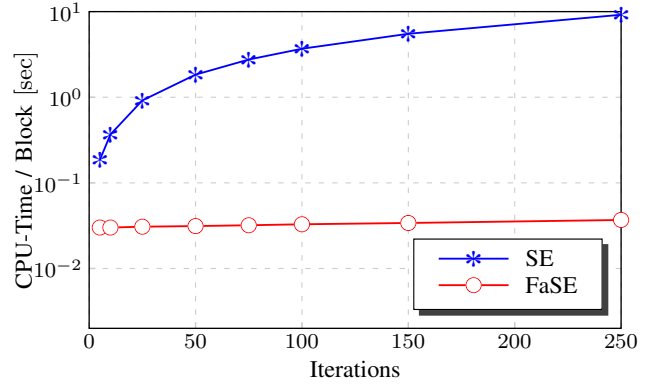

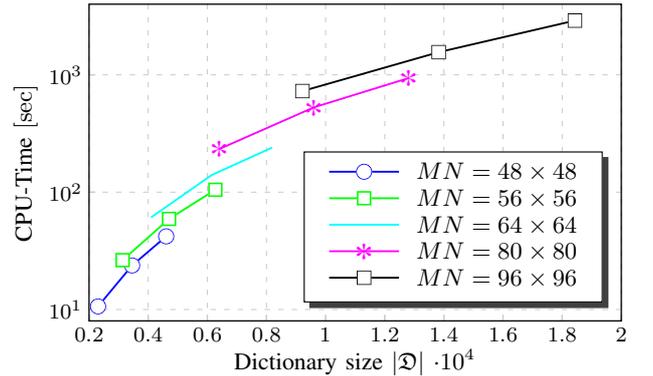
\begin{figure}
	\centering



\providelength{\AxesLineWidth}       \setlength{\AxesLineWidth}{0.5pt}
\providelength{\GridLineWidth}       \setlength{\GridLineWidth}{0.4pt}
\providelength{\GridLineDotSep}      \setlength{\GridLineDotSep}{0.4pt}
\providelength{\MinorGridLineWidth}  \setlength{\MinorGridLineWidth}{0.4pt}
\providelength{\MinorGridLineDotSep} \setlength{\MinorGridLineDotSep}{0.8pt}
\providelength{\plotwidth}           \setlength{\plotwidth}{7cm} 
\providelength{\LineWidth}           \setlength{\LineWidth}{0.7pt}
\providelength{\MarkerSize}          \setlength{\MarkerSize}{6pt}
\newrgbcolor{GridColor}{0.8 0.8 0.8}

\psset{xunit=0.555556\plotwidth,yunit=0.222307\plotwidth}
\begin{pspicture}(-0.057143,0.421131)(2.000000,3.602060)


\psline[linestyle=dashed,dash=2pt 3pt,dotsep=\GridLineDotSep,linewidth=\GridLineWidth,linecolor=GridColor](0.200000,0.903090)(0.200000,3.602060)
\psline[linestyle=dashed,dash=2pt 3pt,dotsep=\GridLineDotSep,linewidth=\GridLineWidth,linecolor=GridColor](0.400000,0.903090)(0.400000,3.602060)
\psline[linestyle=dashed,dash=2pt 3pt,dotsep=\GridLineDotSep,linewidth=\GridLineWidth,linecolor=GridColor](0.600000,0.903090)(0.600000,3.602060)
\psline[linestyle=dashed,dash=2pt 3pt,dotsep=\GridLineDotSep,linewidth=\GridLineWidth,linecolor=GridColor](0.800000,0.903090)(0.800000,3.602060)
\psline[linestyle=dashed,dash=2pt 3pt,dotsep=\GridLineDotSep,linewidth=\GridLineWidth,linecolor=GridColor](1.000000,0.903090)(1.000000,3.602060)
\psline[linestyle=dashed,dash=2pt 3pt,dotsep=\GridLineDotSep,linewidth=\GridLineWidth,linecolor=GridColor](1.200000,0.903090)(1.200000,3.602060)
\psline[linestyle=dashed,dash=2pt 3pt,dotsep=\GridLineDotSep,linewidth=\GridLineWidth,linecolor=GridColor](1.400000,0.903090)(1.400000,3.602060)
\psline[linestyle=dashed,dash=2pt 3pt,dotsep=\GridLineDotSep,linewidth=\GridLineWidth,linecolor=GridColor](1.600000,0.903090)(1.600000,3.602060)
\psline[linestyle=dashed,dash=2pt 3pt,dotsep=\GridLineDotSep,linewidth=\GridLineWidth,linecolor=GridColor](1.800000,0.903090)(1.800000,3.602060)
\psline[linestyle=dashed,dash=2pt 3pt,dotsep=\GridLineDotSep,linewidth=\GridLineWidth,linecolor=GridColor](2.000000,0.903090)(2.000000,3.602060)
\psline[linestyle=dashed,dash=2pt 3pt,dotsep=\GridLineDotSep,linewidth=\GridLineWidth,linecolor=GridColor](0.200000,1.000000)(2.000000,1.000000)
\psline[linestyle=dashed,dash=2pt 3pt,dotsep=\GridLineDotSep,linewidth=\GridLineWidth,linecolor=GridColor](0.200000,2.000000)(2.000000,2.000000)
\psline[linestyle=dashed,dash=2pt 3pt,dotsep=\GridLineDotSep,linewidth=\GridLineWidth,linecolor=GridColor](0.200000,3.000000)(2.000000,3.000000)

\psline[linewidth=\AxesLineWidth,linecolor=GridColor](0.200000,0.903090)(0.200000,0.957069)
\psline[linewidth=\AxesLineWidth,linecolor=GridColor](0.400000,0.903090)(0.400000,0.957069)
\psline[linewidth=\AxesLineWidth,linecolor=GridColor](0.600000,0.903090)(0.600000,0.957069)
\psline[linewidth=\AxesLineWidth,linecolor=GridColor](0.800000,0.903090)(0.800000,0.957069)
\psline[linewidth=\AxesLineWidth,linecolor=GridColor](1.000000,0.903090)(1.000000,0.957069)
\psline[linewidth=\AxesLineWidth,linecolor=GridColor](1.200000,0.903090)(1.200000,0.957069)
\psline[linewidth=\AxesLineWidth,linecolor=GridColor](1.400000,0.903090)(1.400000,0.957069)
\psline[linewidth=\AxesLineWidth,linecolor=GridColor](1.600000,0.903090)(1.600000,0.957069)
\psline[linewidth=\AxesLineWidth,linecolor=GridColor](1.800000,0.903090)(1.800000,0.957069)
\psline[linewidth=\AxesLineWidth,linecolor=GridColor](2.000000,0.903090)(2.000000,0.957069)
\psline[linewidth=\AxesLineWidth,linecolor=GridColor](0.200000,1.000000)(0.221600,1.000000)
\psline[linewidth=\AxesLineWidth,linecolor=GridColor](0.200000,2.000000)(0.221600,2.000000)
\psline[linewidth=\AxesLineWidth,linecolor=GridColor](0.200000,3.000000)(0.221600,3.000000)


{ \footnotesize 
\rput[t](0.200000,0.849111){$0.2$}
\rput[t](0.400000,0.849111){$0.4$}
\rput[t](0.600000,0.849111){$0.6$}
\rput[t](0.800000,0.849111){$0.8$}
\rput[t](1.000000,0.849111){$1$}
\rput[t](1.200000,0.849111){$1.2$}
\rput[t](1.400000,0.849111){$1.4$}
\rput[t](1.600000,0.849111){$1.6$}
\rput[t](1.800000,0.849111){$1.8$}
\rput[t](2.000000,0.849111){$2$}
\rput[r](0.178400,1.000000){$10^{1}$}
\rput[r](0.178400,2.000000){$10^{2}$}
\rput[r](0.178400,3.000000){$10^{3}$}
} 

\pspolygon[linewidth=\AxesLineWidth](0.200000,0.903090)(2.000000,0.903090)(2.000000,3.602060)(0.200000,3.602060)(0.200000,0.903090)

{ \small 
\rput[b](1.100000,0.421131){
\begin{tabular}{c}
Dictionary size $\left|\mathfrak{D}\right|$ $\cdot 10^{4}$\\
\end{tabular}
}

\rput[t]{90}(-0.057143,2.252575){
\begin{tabular}{c}
CPU-Time $\left[\mathrm{sec}\right]$\\
\end{tabular}
}
} 

\newrgbcolor{color933.0122}{0  0  1}
\savedata{\mydata}[{
{0.230400,1.026661},{0.345600,1.375565},{0.460800,1.623609}
}]
\dataplot[plotstyle=line,showpoints=true,dotstyle=o,dotsize=\MarkerSize,linestyle=solid,linewidth=\LineWidth,linecolor=color933.0122]{\mydata}

\newrgbcolor{color934.0117}{0  1  0}
\savedata{\mydata}[{
{0.313600,1.420815},{0.470400,1.772083},{0.627200,2.020617}
}]
\dataplot[plotstyle=line,showpoints=true,dotstyle=Bsquare,dotsize=\MarkerSize,linestyle=solid,linewidth=\LineWidth,linecolor=color934.0117]{\mydata}

\newrgbcolor{color935.0117}{0  1  1}
\savedata{\mydata}[{
{0.409600,1.784860},{0.614400,2.145647},{0.819200,2.379810}
}]
\dataplot[plotstyle=line,showpoints=true,dotstyle=+,dotsize=\MarkerSize,linestyle=solid,linewidth=\LineWidth,linecolor=color935.0117]{\mydata}

\newrgbcolor{color936.0117}{1  0  1}
\savedata{\mydata}[{
{0.640000,2.367239},{0.960000,2.722600},{1.280000,2.971903}
}]
\dataplot[plotstyle=line,showpoints=true,dotstyle=asterisk,dotsize=\MarkerSize,linestyle=solid,linewidth=\LineWidth,linecolor=color936.0117]{\mydata}

\newrgbcolor{color937.0117}{0  0  0}
\savedata{\mydata}[{
{0.921600,2.861482},{1.382400,3.192952},{1.843200,3.462364}
}]
\dataplot[plotstyle=line,showpoints=true,dotstyle=square,dotsize=\MarkerSize,linestyle=solid,linewidth=\LineWidth,linecolor=color937.0117]{\mydata}

{ \small 
\rput[br](1.956800,1.011049){%
\psshadowbox[framesep=0pt,linewidth=\AxesLineWidth]{\psframebox*{\begin{tabular}{l}
\Rnode{a1}{\hspace*{0.0ex}} \hspace*{0.7cm} \Rnode{a2}{~~$MN=48\times 48$} \\
\Rnode{a3}{\hspace*{0.0ex}} \hspace*{0.7cm} \Rnode{a4}{~~$MN=56\times 56$} \\
\Rnode{a5}{\hspace*{0.0ex}} \hspace*{0.7cm} \Rnode{a6}{~~$MN=64\times 64$} \\
\Rnode{a7}{\hspace*{0.0ex}} \hspace*{0.7cm} \Rnode{a8}{~~$MN=80\times 80$} \\
\Rnode{a9}{\hspace*{0.0ex}} \hspace*{0.7cm} \Rnode{a10}{~~$MN=96\times 96$} \\
\end{tabular}}
\ncline[linestyle=solid,linewidth=\LineWidth,linecolor=color933.0122]{a1}{a2} \ncput{\psdot[dotstyle=o,dotsize=\MarkerSize,linecolor=color933.0122]}
\ncline[linestyle=solid,linewidth=\LineWidth,linecolor=color934.0117]{a3}{a4} \ncput{\psdot[dotstyle=Bsquare,dotsize=\MarkerSize,linecolor=color934.0117]}
\ncline[linestyle=solid,linewidth=\LineWidth,linecolor=color935.0117]{a5}{a6} \ncput{\psdot[dotstyle=+,dotsize=\MarkerSize,linecolor=color935.0117]}
\ncline[linestyle=solid,linewidth=\LineWidth,linecolor=color936.0117]{a7}{a8} \ncput{\psdot[dotstyle=asterisk,dotsize=\MarkerSize,linecolor=color936.0117]}
\ncline[linestyle=solid,linewidth=\LineWidth,linecolor=color937.0117]{a9}{a10} \ncput{\psdot[dotstyle=square,dotsize=\MarkerSize,linecolor=color937.0117]}
}%
}%
} 

\end{pspicture}%
\vspace{-2mm}
	\caption{Processing time for generating tables over dictionary size $\left|\mathfrak{D}\right|$ for extrapolation areas of different sizes $MN$.\vspace{-3mm}}
	\label{fig:table_generation}
\end{figure}

In order to support the complexity evaluation from the previous section, the processing time for SE and FaSE is further examined. The first results presented are for arbitrary two-dimensional basis functions. In this case, only the original SE and the novel FaSE can be used, as transform domain algorithms like FSE cannot deal with arbitrary basis functions. For the runtime evaluation, the model generation has been implemented in C, compiled with gcc 4.3.2 and optimizations -O3, and the simulations have been carried out on an Intel Core2 Quad@$2.83 \punit{GHz}$, equipped with $8 \punit{GB}$ RAM. In order to reduce the influence from the operating system, multiple runs of the simulations have been conducted and the computation has been limited to the usage of only one single core. 

For the simulations, a block of size $16\times 16$ samples is extrapolated from its surrounding samples. Furthermore, different sizes of extrapolation area $\mathcal{L}$ between $48\times 48$ and $96\times 96$ samples are regarded. Fig.\ \ref{fig:dct_over_N} shows the extrapolation time per block for different numbers of candidate basis functions and for $250$ iterations performed for model generation. For this plot, the cardinality of the dictionary is selected to be of the same size as the extrapolation area. Thus, $\mathfrak{D}$ varies between $\left|\mathfrak{D}\right|=2304$ basis functions of size $48\times 48$ and $\left|\mathfrak{D}\right|=9216$ basis functions of size $96\times 96$. Comparing the two curves of SE and FaSE one can easily recognize that FaSE is about $250$ times faster than the original SE, independently of the problem size. This is due to the fact, that for FaSE the computationally expensive weighted scalar products only have to be evaluated once, namely prior to the first iteration. In the later iterations, the expensive steps can be avoided by making use of the tabulated values and avoiding the update of the residual. For these evaluations, the calculation time for generating the tabulated values is not considered, as they only have to be computed once and can be stored. The very high computational cost of the weighted scalar products can also be recognized by regarding Fig.\ \ref{fig:dct_over_iter} that shows the extrapolation time per block over iterations for an extrapolation scenario of size $64\times 64$ samples. Taking into account the logarithmic axis, one can recognize that the processing time per block more or less linearly increases for SE, whereas for FaSE the processing time per block increases only very slowly. The results correspond well to the analytical complexity evaluation and the speed gain of FaSE over SE is of the same magnitude as shown in the previous section. Apparently, Fig.\ \ref{fig:complexity} cannot be directly translated into the processing time shown in Fig.\ \ref{fig:dct_over_iter} since not all regarded operations consume the same processing time and since the analytical evaluation cannot account for optimizations introduced by the compiler. 

Fig.\ \ref{fig:table_generation} shows the processing time for generating the tables for different dictionary sizes $\left|\mathfrak{D}\right|$ and for different sizes of extrapolation area $\mathcal{L}$. Comparing these results with the ones shown in Fig. \ref{fig:dct_over_iter} one can recognize that for an extrapolation area of size $64\times 64$, a dictionary size of $\left|\mathfrak{D}\right|=4096$ and $250$ iterations, the table generation only takes as long as SE would roughly need for extrapolating $6$ blocks. This corresponds well to the theoretical results presented in the analytical evaluation. The discrepancy follows from the fact that different operations consume unequal amounts of processing time while in the analytical evaluation only the absolute number of operations has been counted. 

\begin{table}
\centering
\caption{Average results for extrapolation of $126$ blocks of size $16\times 16$ samples in every image of the Kodak test image database.}
\begin{tabular}{|c|c|c|}
\hline
Algorithm & $\PSNR$  & Processing time per block  \\ \hline
TV \cite{Dahl2010} & $22.40\punit{dB}$ & $0.54\punit{sec}$ \\ \hline
SFG \cite{Li2008} & $23.63\punit{dB}$ & $15.29\punit{sec}$ \\ \hline
SDI \cite{Alkachouh2000} & $21.60\punit{dB}$ & $0.0003\punit{sec}$ \\ \hline
FaSE & $23.82\punit{dB}$ & $0.38\punit{sec}$ \\ \hline
\end{tabular}
\label{tab:extrapolation_quality}\vspace{-3mm}
\end{table}

Since the proposed novel spatial domain solution does not affect the model generation principle of SE, still a very high extrapolation quality can be achieved. Due to the acceleration of the algorithm, now very good extrapolation results can be achieved at a manageable complexity for arbitrary basis functions. To prove this, Table \ref{tab:extrapolation_quality} shows the average extrapolation quality in terms of $\PSNR$ and the processing time for extrapolating $126$ blocks of size $16\times 16$ samples in every image from the Kodak image database. For comparison, the Total Variation Image Reconstruction (TV) algorithm from \cite{Dahl2010}, the patch-based algorithm from \cite{Li2008} that uses Stochastic Factor Graphs (SFG) and the simple but very fast Spatial-Domain Interpolation (SDI) from \cite{Alkachouh2000} are regarded. The comparison has been carried out in MATLAB R2008b, and again only one core of the above mentioned computer has been used. Apparently, FaSE provides the highest extrapolation quality among the considered algorithms only with SFG coming close. But at the same time, it is second fastest algorithm.


\section{Modifications for Transform-Based Basis Function Sets}\label{sec:results_transform} 

As aforementioned, for FaSE the weighted scalar products only have to be evaluated prior to the first iteration. In the case that the regarded basis function set contains a subset of basis functions that emanate from a discrete transform as \mbox{e.\ g.\ } functions of the DCT or the DFT, the explicit evaluation of the weighted scalar products can be simplified by replacing the summation over the product between the weighted signal and the basis function by the corresponding transform coefficient of the weighted signal which can be achieved through a fast transform. To give an example, the idea that the basis function set contains some basis functions which emanate from the DFT will be extended. In this case, a basis function is defined by
\begin{equation}
 \varphi_k\left[m,n\right] = \e^{\j\frac{2\pi}{M}\mu_k m}\e^{\j\frac{2\pi}{N}\eta_k n}
\end{equation}
with vertical frequency $\mu_k$ and horizontal frequency $\eta_k$. Then, the summation from (\ref{eq:scalar0}) can be expressed by the DFT
\begin{equation}
\label{eq:dft_property}
 \sum_{\left(m,n\right)\in\mathcal{L}}\hspace{-3.5mm} s\hspace{-0.5mm}\left[m,n\right] \varphi_k^\ast\hspace{-0.5mm}\left[m,n\right]w\hspace{-0.5mm}\left[m,n\right] \hspace{-1mm}=\hspace{-1mm} \mathrm{DFT}\hspace{-1mm}\left\{s\hspace{-0.5mm}\left[m,n\right]w\hspace{-0.5mm}\left[m,n\right]\right\}|_{\mu_k,\eta_k}
\end{equation}
at frequency $\mu_k,\eta_k$. Thus, the weighted scalar products for many basis functions can be efficiently evaluated simultaneously by making use of fast transforms like the Fast Fourier Transform \cite{Cooley1965} or respectively a fast transform that is appropriate to the regarded basis functions. It has to be noted that the utilization of fast transforms is only reasonable if a large number of transform domain coefficients has to be calculated at the same time. The fast transforms only speed up the parallel calculation of many coefficients. The calculation of just a single coefficient would take as long as the explicit evaluation of the weighted scalar product. The above described property could also be used for speeding up the table generation in (\ref{eq:table_generation}). Regarding again the example of a subset of DFT basis functions, the product between a basis function and a conjugate complex second one is equal to a basis function where the horizontal and vertical frequency results from the difference of the original frequencies:
\begin{eqnarray}
 \nonumber \hspace{-3mm}\varphi_k^\ast\left[m,n\right]\varphi_l\left[m,n\right] \hspace{-3mm} &=& \hspace{-3mm}\e^{-\j\frac{2\pi}{M}\mu_k m}\e^{-\j\frac{2\pi}{N}\eta_k n}\e^{\j\frac{2\pi}{M}\mu_l m}\e^{\j\frac{2\pi}{N}\eta_l n} \\
 \hspace{-3mm}&=&\hspace{-3mm}\e^{\j\frac{2\pi}{M}\left(\mu_l-\mu_k\right) m}\e^{\j\frac{2\pi}{N}\left(\eta_l-\eta_k\right) n}
\end{eqnarray}
Hence, (\ref{eq:table_generation}) can also be expressed by the corresponding coefficients from the DFT. For other transform-based basis function sets similar properties exist.

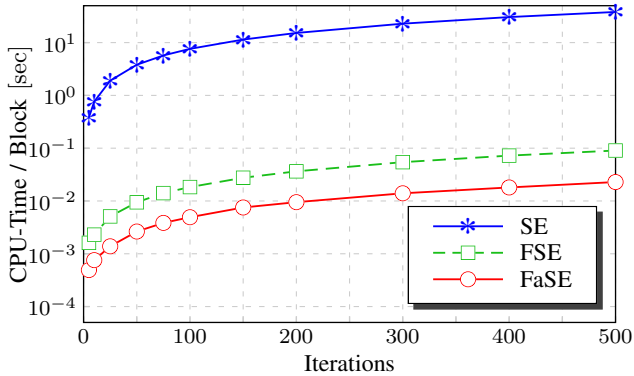
\begin{figure}
	\centering



\providelength{\AxesLineWidth}       \setlength{\AxesLineWidth}{0.5pt}
\providelength{\GridLineWidth}       \setlength{\GridLineWidth}{0.4pt}
\providelength{\GridLineDotSep}      \setlength{\GridLineDotSep}{0.4pt}
\providelength{\MinorGridLineWidth}  \setlength{\MinorGridLineWidth}{0.4pt}
\providelength{\MinorGridLineDotSep} \setlength{\MinorGridLineDotSep}{0.8pt}
\providelength{\plotwidth}           \setlength{\plotwidth}{7cm} 
\providelength{\LineWidth}           \setlength{\LineWidth}{0.7pt}
\providelength{\MarkerSize}          \setlength{\MarkerSize}{6pt}
\newrgbcolor{GridColor}{0.8 0.8 0.8}

\psset{xunit=0.002000\plotwidth,yunit=0.100000\plotwidth}
\begin{pspicture}(-71.428571,-5.372459)(500.000000,1.698970)


\psline[linestyle=dashed,dash=2pt 3pt,dotsep=\GridLineDotSep,linewidth=\GridLineWidth,linecolor=GridColor](0.000000,-4.301030)(0.000000,1.698970)
\psline[linestyle=dashed,dash=2pt 3pt,dotsep=\GridLineDotSep,linewidth=\GridLineWidth,linecolor=GridColor](50.000000,-4.301030)(50.000000,1.698970)
\psline[linestyle=dashed,dash=2pt 3pt,dotsep=\GridLineDotSep,linewidth=\GridLineWidth,linecolor=GridColor](100.000000,-4.301030)(100.000000,1.698970)
\psline[linestyle=dashed,dash=2pt 3pt,dotsep=\GridLineDotSep,linewidth=\GridLineWidth,linecolor=GridColor](150.000000,-4.301030)(150.000000,1.698970)
\psline[linestyle=dashed,dash=2pt 3pt,dotsep=\GridLineDotSep,linewidth=\GridLineWidth,linecolor=GridColor](200.000000,-4.301030)(200.000000,1.698970)
\psline[linestyle=dashed,dash=2pt 3pt,dotsep=\GridLineDotSep,linewidth=\GridLineWidth,linecolor=GridColor](250.000000,-4.301030)(250.000000,1.698970)
\psline[linestyle=dashed,dash=2pt 3pt,dotsep=\GridLineDotSep,linewidth=\GridLineWidth,linecolor=GridColor](300.000000,-4.301030)(300.000000,1.698970)
\psline[linestyle=dashed,dash=2pt 3pt,dotsep=\GridLineDotSep,linewidth=\GridLineWidth,linecolor=GridColor](350.000000,-4.301030)(350.000000,1.698970)
\psline[linestyle=dashed,dash=2pt 3pt,dotsep=\GridLineDotSep,linewidth=\GridLineWidth,linecolor=GridColor](400.000000,-4.301030)(400.000000,1.698970)
\psline[linestyle=dashed,dash=2pt 3pt,dotsep=\GridLineDotSep,linewidth=\GridLineWidth,linecolor=GridColor](450.000000,-4.301030)(450.000000,1.698970)
\psline[linestyle=dashed,dash=2pt 3pt,dotsep=\GridLineDotSep,linewidth=\GridLineWidth,linecolor=GridColor](500.000000,-4.301030)(500.000000,1.698970)
\psline[linestyle=dashed,dash=2pt 3pt,dotsep=\GridLineDotSep,linewidth=\GridLineWidth,linecolor=GridColor](0.000000,-4.000000)(500.000000,-4.000000)
\psline[linestyle=dashed,dash=2pt 3pt,dotsep=\GridLineDotSep,linewidth=\GridLineWidth,linecolor=GridColor](0.000000,-3.000000)(500.000000,-3.000000)
\psline[linestyle=dashed,dash=2pt 3pt,dotsep=\GridLineDotSep,linewidth=\GridLineWidth,linecolor=GridColor](0.000000,-2.000000)(500.000000,-2.000000)
\psline[linestyle=dashed,dash=2pt 3pt,dotsep=\GridLineDotSep,linewidth=\GridLineWidth,linecolor=GridColor](0.000000,-1.000000)(500.000000,-1.000000)
\psline[linestyle=dashed,dash=2pt 3pt,dotsep=\GridLineDotSep,linewidth=\GridLineWidth,linecolor=GridColor](0.000000,0.000000)(500.000000,0.000000)
\psline[linestyle=dashed,dash=2pt 3pt,dotsep=\GridLineDotSep,linewidth=\GridLineWidth,linecolor=GridColor](0.000000,1.000000)(500.000000,1.000000)

\psline[linewidth=\AxesLineWidth,linecolor=GridColor](0.000000,-4.301030)(0.000000,-4.181030)
\psline[linewidth=\AxesLineWidth,linecolor=GridColor](50.000000,-4.301030)(50.000000,-4.181030)
\psline[linewidth=\AxesLineWidth,linecolor=GridColor](100.000000,-4.301030)(100.000000,-4.181030)
\psline[linewidth=\AxesLineWidth,linecolor=GridColor](150.000000,-4.301030)(150.000000,-4.181030)
\psline[linewidth=\AxesLineWidth,linecolor=GridColor](200.000000,-4.301030)(200.000000,-4.181030)
\psline[linewidth=\AxesLineWidth,linecolor=GridColor](250.000000,-4.301030)(250.000000,-4.181030)
\psline[linewidth=\AxesLineWidth,linecolor=GridColor](300.000000,-4.301030)(300.000000,-4.181030)
\psline[linewidth=\AxesLineWidth,linecolor=GridColor](350.000000,-4.301030)(350.000000,-4.181030)
\psline[linewidth=\AxesLineWidth,linecolor=GridColor](400.000000,-4.301030)(400.000000,-4.181030)
\psline[linewidth=\AxesLineWidth,linecolor=GridColor](450.000000,-4.301030)(450.000000,-4.181030)
\psline[linewidth=\AxesLineWidth,linecolor=GridColor](500.000000,-4.301030)(500.000000,-4.181030)
\psline[linewidth=\AxesLineWidth,linecolor=GridColor](0.000000,-4.000000)(6.000000,-4.000000)
\psline[linewidth=\AxesLineWidth,linecolor=GridColor](0.000000,-3.000000)(6.000000,-3.000000)
\psline[linewidth=\AxesLineWidth,linecolor=GridColor](0.000000,-2.000000)(6.000000,-2.000000)
\psline[linewidth=\AxesLineWidth,linecolor=GridColor](0.000000,-1.000000)(6.000000,-1.000000)
\psline[linewidth=\AxesLineWidth,linecolor=GridColor](0.000000,0.000000)(6.000000,0.000000)
\psline[linewidth=\AxesLineWidth,linecolor=GridColor](0.000000,1.000000)(6.000000,1.000000)

{ \footnotesize 
\rput[t](0.000000,-4.421030){$0$}
\rput[t](100.000000,-4.421030){$100$}
\rput[t](200.000000,-4.421030){$200$}
\rput[t](300.000000,-4.421030){$300$}
\rput[t](400.000000,-4.421030){$400$}
\rput[t](500.000000,-4.421030){$500$}
\rput[r](-6.000000,-4.000000){$10^{-4}$}
\rput[r](-6.000000,-3.000000){$10^{-3}$}
\rput[r](-6.000000,-2.000000){$10^{-2}$}
\rput[r](-6.000000,-1.000000){$10^{-1}$}
\rput[r](-6.000000,0.000000){$10^{0}$}
\rput[r](-6.000000,1.000000){$10^{1}$}
} 

\pspolygon[linewidth=\AxesLineWidth](0.000000,-4.301030)(500.000000,-4.301030)(500.000000,1.698970)(0.000000,1.698970)(0.000000,-4.301030)

{ \small 
\rput[b](250.000000,-5.372459){
\begin{tabular}{c}
Iterations\\
\end{tabular}
}

\rput[t]{90}(-71.428571,-1.301030){
\begin{tabular}{c}
CPU-Time / Block $\left[\mathrm{sec}\right]$\\
\end{tabular}
}
} 

\newrgbcolor{color693.0138}{1  0  0}
\savedata{\mydata}[{
{5.000000,-3.308515},{10.000000,-3.120536},{25.000000,-2.858752},{50.000000,-2.578886},{75.000000,-2.415219},
{100.000000,-2.307742},{150.000000,-2.123404},{200.000000,-2.022633},{300.000000,-1.854586},{400.000000,-1.743852},
{500.000000,-1.641279}
}]
\dataplot[plotstyle=line,showpoints=true,dotstyle=o,dotsize=\MarkerSize,linestyle=solid,linewidth=\LineWidth,linecolor=color693.0138]{\mydata}

\newrgbcolor{color695.0133}{0  0  1}
\savedata{\mydata}[{
{5.000000,-0.418088},{10.000000,-0.117829},{25.000000,0.279404},{50.000000,0.580564},{75.000000,0.756824},
{100.000000,0.881618},{150.000000,1.057635},{200.000000,1.182746},{300.000000,1.358616},{400.000000,1.483607},
{500.000000,1.580572}
}]
\dataplot[plotstyle=line,showpoints=true,dotstyle=asterisk,dotsize=\MarkerSize,linestyle=solid,linewidth=\LineWidth,linecolor=color695.0133]{\mydata}

\newrgbcolor{color696.0133}{0.1  0.75  0.1}
\savedata{\mydata}[{
{5.000000,-2.792821},{10.000000,-2.637569},{25.000000,-2.292554},{50.000000,-2.025426},{75.000000,-1.850992},
{100.000000,-1.736654},{150.000000,-1.560042},{200.000000,-1.438605},{300.000000,-1.264783},{400.000000,-1.140785},
{500.000000,-1.044255}
}]
\dataplot[plotstyle=line,showpoints=true,dotstyle=square,dotsize=\MarkerSize,linestyle=dashed,linewidth=\LineWidth,linecolor=color696.0133]{\mydata}

{ \small 
\rput[br](488.000000,-4.061030){%
\psshadowbox[framesep=0pt,linewidth=\AxesLineWidth]{\psframebox*{\begin{tabular}{l}
\Rnode{a5}{\hspace*{0.0ex}} \hspace*{0.7cm} \Rnode{a6}{~~SE} \\
\Rnode{a7}{\hspace*{0.0ex}} \hspace*{0.7cm} \Rnode{a8}{~~FSE} \\
\Rnode{a1}{\hspace*{0.0ex}} \hspace*{0.7cm} \Rnode{a2}{~~FaSE} \\
\end{tabular}}
\ncline[linestyle=solid,linewidth=\LineWidth,linecolor=color693.0138]{a1}{a2} \ncput{\psdot[dotstyle=o,dotsize=\MarkerSize,linecolor=color693.0138]}
\ncline[linestyle=solid,linewidth=\LineWidth,linecolor=color695.0133]{a5}{a6} \ncput{\psdot[dotstyle=asterisk,dotsize=\MarkerSize,linecolor=color695.0133]}
\ncline[linestyle=dashed,linewidth=\LineWidth,linecolor=color696.0133]{a7}{a8} \ncput{\psdot[dotstyle=square,dotsize=\MarkerSize,linecolor=color696.0133]}
}%
}%
} 

\end{pspicture}%
\vspace{-2mm}
	\caption{Processing time over iterations for 2D model generation with DFT basis functions, $\left|\mathfrak{D}\right|=4096$.\vspace{-3mm}}
	\label{fig:dft_over_iter}
\end{figure}

In addition to the results for arbitrary basis functions shown in Section \ref{sec:results_arbitrary}, the performance of FaSE and SE is compared to a transform domain algorithm. For this, FSE is regarded that utilizes Fourier functions for extrapolation. Here the circumstance has to be considered, that, as described in \cite{Kaup2005}, FSE does not generate a complex-valued model. FSE selects in every iteration step one basis function and its corresponding conjugate complex one, in such a way that the model always is real-valued. Hence, in most cases two basis functions are selected in an iteration, with the exception of the real-valued constant basis function and the function with the highest possible alternation. Thus, the number of iterations has to be doubled for SE and FaSE for a fair comparison as they select only one basis function per iteration. Fig.\ \ref{fig:dft_over_iter} shows the processing time per block for the different approaches with $\left|\mathfrak{D}\right|= 4096$ Fourier basis functions of size $64\times 64$. For these simulations, the initial scalar products for FaSE are expressed by the transform coefficients according to (\ref{eq:dft_property}). Although FaSE needs twice the number of iterations as FSE for generating the model, it is still significantly faster than FSE and furthermore several magnitudes faster than the original spatial domain SE. 

Taking all the results from the two previous sections into account, the following recommendations can be given. In the case that the Selective Extrapolation is carried out with Fourier basis functions or other basis function sets that are based on a discrete transform, one can decide either to use a transform domain algorithm or the novel FaSE. If always the same extrapolation scenario is considered, the tables only have to calculated once and the time gain of FaSE prevails, otherwise the transform domain algorithm is the better choice as no calculation of the tables is necessary. If the extrapolation process is carried out with basis functions for which no transform domain implementation is possible, FaSE should be preferred over the original SE. FaSE is able to efficiently trade computational complexity versus memory consumption as the expensive operations only have to be carried out once. Thus, the actual iterations for generating the model become very simple and very fast.


\section{Conclusion} \label{sec:conclusion} 

Within the scope of this contribution, we presented Fast Selective Extrapolation for image and video signal extrapolation. For this, Selective Extrapolation, a powerful signal extrapolation algorithm has been reviewed and its most complex parts have been identified. The novel algorithm behaves mathematically identical to the original algorithm but is able to outspeed the original algorithm by several decades by effectively trading memory consumption versus processing time. Furthermore, the novel algorithm is able to outperform existent fast transform domain extrapolation algorithms which are even limited to certain basis function sets. With that, it opens the door for further research on carrying out the extrapolation with different basis function sets. Up to now, the extrapolation only has been computationally manageable for special basis function sets that are based on discrete transforms. But by using Fast Selective Extrapolation, the extrapolation can be carried out for arbitrary basis functions which may even be only numerically defined. This ability allows for further research on extrapolation with signal adapted basis functions, obtained through the Karhunen-Lo{\`e}ve Transform \cite{Karhunen1947, Loeve1948}, which has not been computationally feasible up to now.

Although the algorithm has been introduced only for two-dimensional data sets, it can be extended straightforwardly to three dimensions by making use of the ideas from \cite{Meisinger2007} and four dimensions by using \cite{Fecker2008}. There, a three-dimensional or respectively a four-dimensional model is generated in the same way as described above for two dimensions.



\end{document}